\documentclass[aps,prl,twocolumn,showpacs, notitlepage]{revtex4-1}
\usepackage{graphicx}
\usepackage{natbib}
\usepackage[utf8]{inputenc}
\usepackage{epstopdf}
\setcitestyle{super}
\usepackage{color}
\usepackage[normalem]{ulem}
\usepackage{pdfpages}
\makeatletter
\AtBeginDocument{\let\LS@rot\@undefined}
\makeatother

\usepackage{hyphenat}
\begin{document}

\preprint{APS/123-QED}

\title{Bulk superconductivity and role of fluctuations in the iron-based superconductor FeSe at high pressures}
\author{Elena Gati$^{1,2}$}
\author{Anna E. B\"ohmer$^{1,2,\star}$}
\author{Sergey L. Bud'ko$^{1,2}$}
\author{Paul C. Canfield$^{1,2}$}

\address{$^{1}$ Ames Laboratory, US Department of Energy, Iowa State University, Ames,
Iowa 50011, USA}
\address{$^{2}$ Department of Physics and Astronomy, Iowa State University, Ames, Iowa 50011, USA}
\address{$^\star$ current address: Institute for Solid State Physics, Karlsruhe Institute of Technology, 76021 Karlsruhe, Germany}

\date{\today}

\begin{abstract}
The iron-based superconductor FeSe offers a unique possibility to study the interplay of superconductivity with purely nematic as well magnetic-nematic order by pressure ($p$) tuning. By measuring specific heat under $p$ up to 2.36\,GPa, we study the multiple phases in FeSe using a thermodynamic probe. We conclude that superconductivity is bulk across the entire $p$ range and competes with magnetism. In addition, whenever magnetism is present, fluctuations exist over a wide temperature range above both the bulk superconducting and the magnetic transitions. Whereas the magnetic fluctuations are likely temporal, the superconducting fluctuations may be either temporal or spatial. These observations highlight similarities between FeSe and underdoped cuprate superconductors.
\end{abstract}

\pacs{xxx}

\maketitle

FeSe is considered to be an exceptional member \cite{Boehmer17,Coldea18} of the family of iron (Fe)-based superconductors \cite{Paglione10,Johnston10,Stewart11,Hosono15,Canfield10} for various reasons. First, FeSe is the structurally simplest of all members. It superconducts \cite{Hsu08} below a critical temperature $T_c\,\approx\,8\,$K and $T_c$ can be significantly enhanced in thin films \cite{Ge15,Sadovskii16,Wang17,Huang17} and intercalated FeSe \cite{Burrard-Lucas13} or by pressure ($p$) \cite{Mizuguchi08,Medvedev09,Margadonna09,Garbarino09,Masaki09,Okabe10}. Second, FeSe undergoes a structural transition \cite{Hsu08,Margadonna08,McQueen09} from a tetragonal to an orthorhombic state at $T_s\,\approx\,90\,$K at ambient $p$ which was shown to be nematic \cite{Boehmer15,Watson15,Tanatar16,Baek15}, i.e., driven by electronic degrees of freedom. In contrast to other Fe-based superconductors \cite{Fernandes14}, the nematic transition in FeSe is not accompanied or closely followed by an antiferromagnetic transition \cite{Bendele10,McQueen09}. Thus, it was suggested that FeSe represents an ideal platform to study a purely nematic phase and its interrelation with superconductivity \cite{Boehmer17}. Third, FeSe was found to be characterized by strong electronic correlations \cite{Yi15} leading to a small Fermi energy \cite{Coldea18} which is comparable in size to the superconducting gap. This has recently raised the question whether FeSe is located deep in the crossover regime between weak-coupling BCS to strong-coupling BEC superconductivity \cite{Kasahara14,Kasahara16,Watashige17,Rinott17,Hanaguri19,Lubashevsky12}. The latter is characterized by superconducting fluctuations over a wide temperature ($T$) range above $T_c$.

\begin{center}
	\begin{figure}
	\includegraphics[width=\columnwidth]{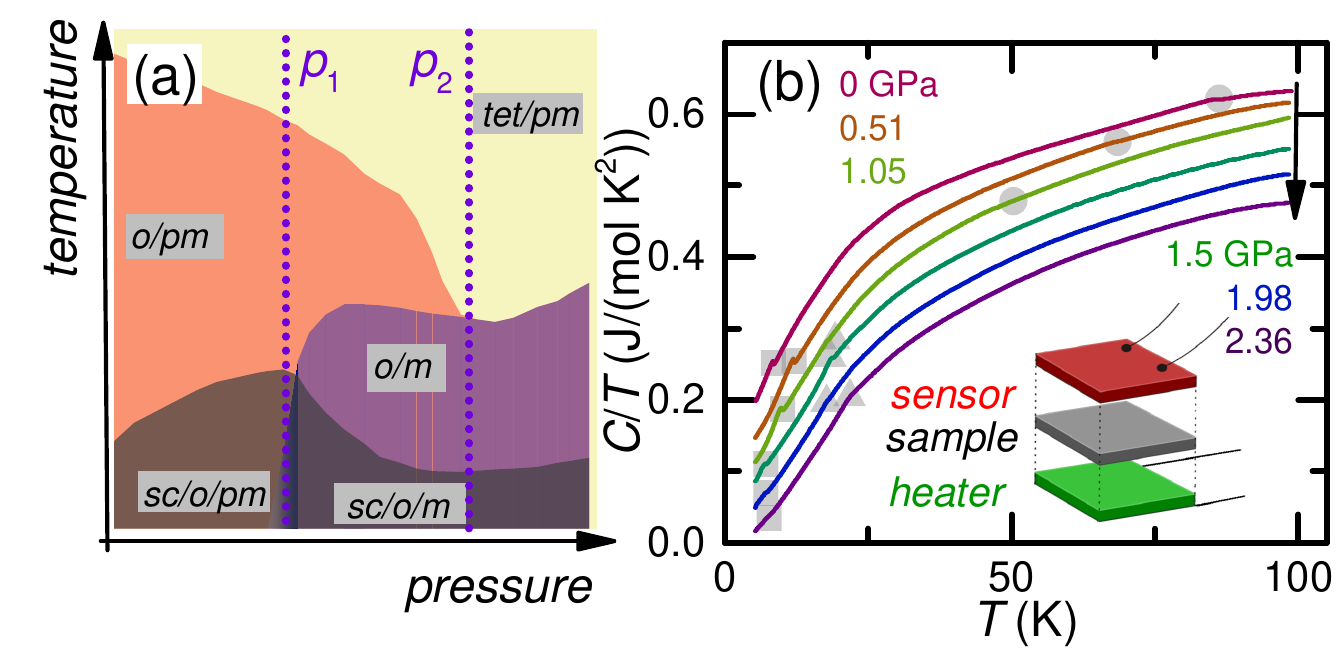} 
	\caption{(a) Schematic temperature-pressure phase diagram of FeSe, showing the extent of tetragonal (tet), orthorhombic (o), paramagnetic (pm), magnetic (m) and superconducting (sc) states and the two characteristic pressures $p_1$ and $p_2$ (see main text); (b) Selected specific heat data sets, $C/T$ vs. $T$, at different pressures. Light grey regions indicate the position of the various anomalies detected by $C/T$, related to the structural (circles), the superconduting (squares) and the magnetic transition (triangles). The inset illustrates schematically the measurement configuration\cite{Gati19} to measure the specific heat under $p$.}
	\label{fig:schematics}
	\end{figure}		
	\end{center}

The extent to which the properties of FeSe are comparable to those of other Fe-based superconductors has been strongly debated over the years \cite{Boehmer17}. In this regard, the study of the $T$-$p$ phase diagram (see Fig.\,\ref{fig:schematics} (a)) yielded important new insights \cite{Bendele10,Kothapalli16,Wang16,Wiecki17,Kaluarachchi16,Miyoshi14,Knoener15,Terashima15,Sun16,Bendele12,Khasanov18,Boehmer18,Svitlyk17,Lebert18} (see Fig.\,S1). Above a characteristic pressure $p_1$, bulk magnetic order \cite{Bendele10,Bendele12}, which is likely stripe-type antiferromagnetic order \cite{Wang16,Khasanov17,Kothapalli16}, was observed at the magnetic transition temperature $T_M\,<\,T_s$ (i.e., the magnetic-nematic state). At even higher pressures, above a second characteristic pressure $p_2$, the magnetic-nematic ground state was found to be stabilized through a simultaneous first-order transition  with $T_s\,=\,T_M$ \cite{Kothapalli16,Wang16,Khasanov18}. This demonstrated that the phase diagram of FeSe at higher $p$ shows the same generic features in terms of the magnetic and structural transitions as other Fe-based superconductors, i.e., two subsequent, second-order phase transitions with $T_s\,>\,T_M$ that can be tuned to a simultaneous first-order transition ($T_s\,=\,T_M$) \cite{Kothapalli16,Wang16,Khasanov18}. However, whether the purely nematic state at low pressures fits into this universal picture, is still a subject of debates \cite{Fernandes16,Glasbrenner15,Onari16,Yamakawa16,Chubukov15,Yu15,Wang15}.

With respect to the superconductivity of FeSe under pressure, there is an ongoing discussion about its nature. It was proposed early on that superconductivity exists over a wide $p$ range, i.e., in the purely nematic ($p\,<\,p_1$), but also in the magnetic-nematic $p$ range ($p\,>\,p_1$). In the latter regime, the simultaneous enhancement of $T_c$ and $T_M$ raised the idea of cooperative promotion of superconductivity and magnetism \citep{Bendele12,Chen19}, contrary to other Fe-based superconductors. However, this scenario has not be substantiated to date, since microscopic probes, such as NMR \cite{Wang16}, failed to detect any signature of superconductivity in the magnetic-nematic state for $p\,>\,p_2$. This has therefore even led to the question whether bulk superconductivity exists in FeSe for $p\,>\,p_2$ \cite{Wang16,Yip17}.

By studying the specific heat ($C$) under $p$ of a single crystal\cite{Boehmer16} of FeSe up to 2.36\,GPa, we determine the full thermodynamic $T$-$p$ phase diagram of FeSe. We are therefore able to address various open issues related to superconductivity: our results confirm the bulk nature of superconductivity over the full $p$ range investigated, in particular also in the magnetic-nematic state for $p\,>\,p_2$. In this regime, our data suggest a competition of superconductivity and magnetism in FeSe. Even further, we argue that superconducting and magnetic fluctuations of temporal and/or spatial nature exist in FeSe at high $p$ over a wide range of temperatures above the respective bulk transition temperatures. These results therefore put FeSe in close similarity to the strongly correlated cuprate superconductors. \\

The specific heat of a vapor grown FeSe single crystal \cite{Boehmer16} was measured with an ac-technique (see Fig.\,\ref{fig:schematics}\,(b)) inside a liquid-medium piston-cylinder pressure cell in a home-built setup \cite{Gati19} (for more details, see SI).

	\begin{figure}
	\includegraphics[width=\columnwidth]{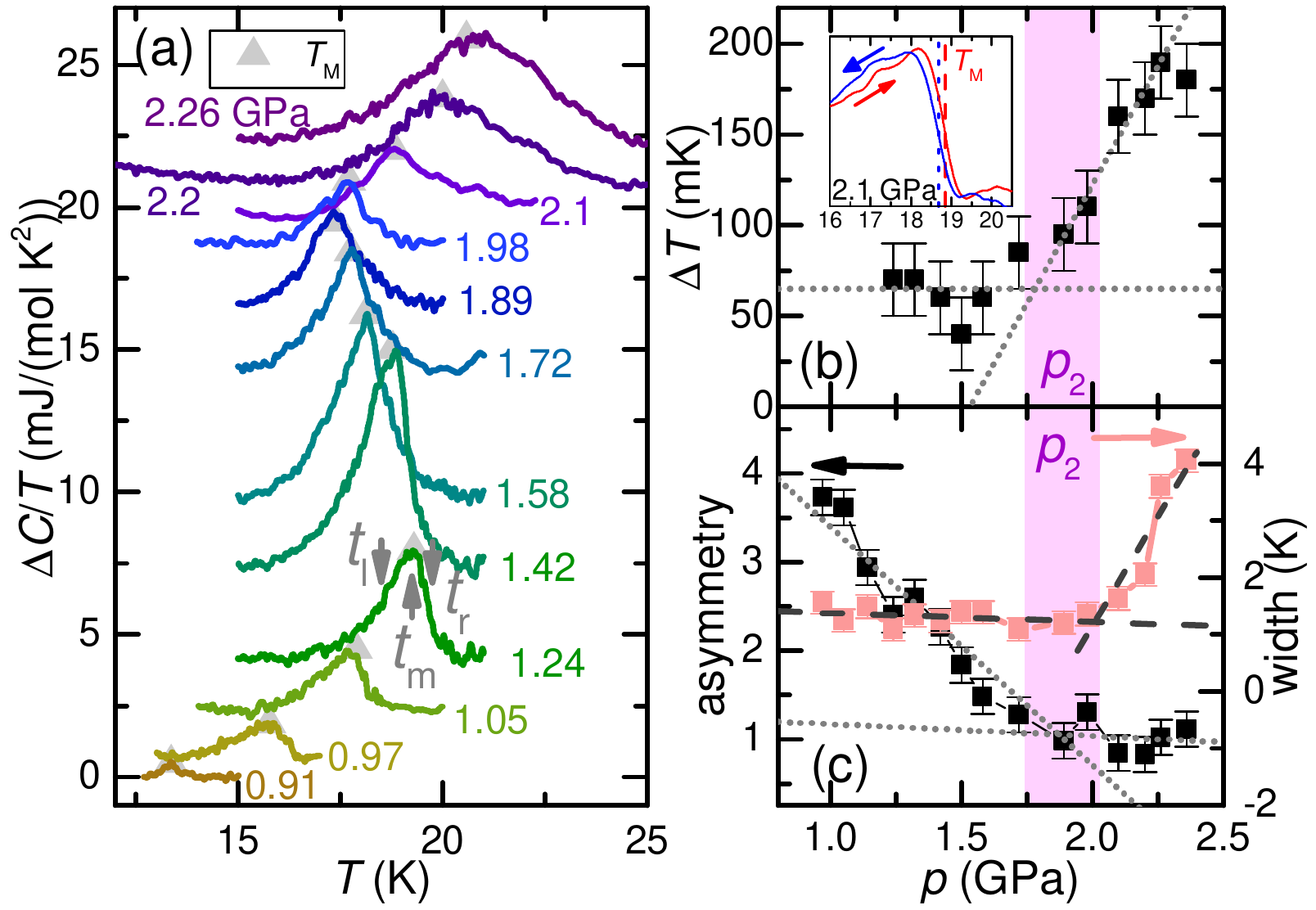} 
	\caption{(a) Specific heat anomaly of the magnetic transition at $T_M$, $\Delta C/T$, which is present for $p\,\ge\,0.91\,$GPa ($\sim p_1$) and obtained by subtracting a background from $C/T$ data. Data are offset for clarity. Faint, grey triangles indicate the position of $T_M$ in each data set. $t_l$, $t_m$ and $t_r$ are used to estimate asymmetry and width of the specific heat peak; (b) Hysteresis $\Delta T$ of $T_M$  between warming and cooling. Inset shows d($C/T$)/d$T$ at 2.1\,GPa upon warming and cooling; (c) Asymmetry (left axis) and width (right axis) of the specific heat peak at $T_M$. Dashed and dotted lines are guides to the eye, the purple bar indicates the position of the critical pressure range $p_2$.}
	\label{fig:data-magnetic-structural}
	\end{figure}	
	
First, we focus on the $C$ data close to the structural and magnetic transitions at $T_s$ and $T_M$, respectively, in FeSe under $p$, as shown in Fig.\,\ref{fig:schematics} (b) and \ref{fig:data-magnetic-structural} (and in Figs.\,S2-S7) to determine the characteristic pressures $p_1$ and $p_2$ from our experiment. $T_s$ is monotonically suppressed with increasing $p$ until it becomes indiscernible above 1.32\,GPa (see Figs.\,\ref{fig:schematics} (b) and S3). Magnetic ordering is observed in our data for $p\,\ge\,0.91\,$GPa (see Fig.\,\ref{fig:data-magnetic-structural} (a) and Fig.\,S5 for low $p$ data). This therefore defines $p_1$ in the $T$-$p$ phase diagram of FeSe (0.84\,GPa$\,\le\,p_1\,\le\,$0.91\,GPa). 

Upon increasing $p$, $T_M$ first increases steeply up to $\approx\,$1.2\,GPa, then shows a slight reduction up to $\approx\,$1.9\,GPa and then increases quickly for higher pressures. At the same time, the specific heat anomaly at $T_M$ (see Fig.\,\ref{fig:data-magnetic-structural}\,(a)) evolves from a step-like shape, characteristic for second-order phase transitions at lower $p$, to a symmetric peak at higher $p$, which might be the result of a slightly broadened singularity of a first-order transition. This observation is therefore consistent with the picture \cite{Kothapalli16,Wang16} that the magnetic transition becomes first order close to where it merges with the structural transition. To define the characteristic pressure $p_2$ at which the character of the magnetic transition changes, we follow three complimentary approaches. This includes measurements of the thermal hysteresis (see Fig.\,\ref{fig:data-magnetic-structural}\,(b) and Fig.\,S7) and an analysis of the asymmetry and the width of the specific heat peak (see Fig.\,\ref{fig:data-magnetic-structural}\,(c)). We define the asymmetry as $\frac{t_r-t_m}{t_m-t_l}$, with $t_m$ ($t_r$ and $t_l$) being the temperatures at which the specific heat anomaly exhibits its maximum value (50\,$\%$ of the maximum value) and the width as $t_r-t_l$. All together, all three quantities exhibit a sudden change at $p_2\,=\,(1.87\,\pm\,0.10)\,$GPa.
	
		\begin{center}
		\begin{figure}[h!]
	\includegraphics[width=\columnwidth]{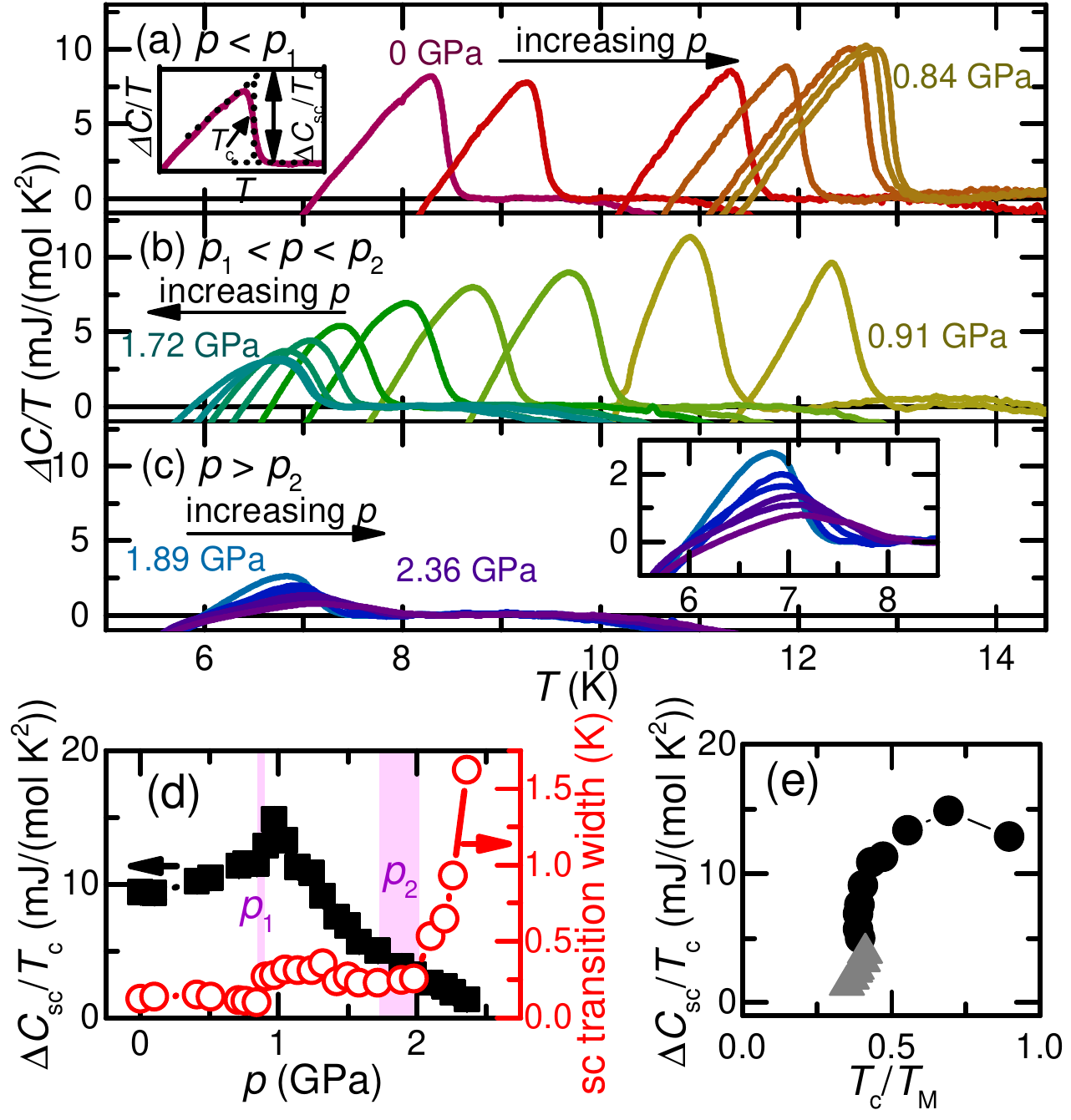} 
	\caption{(a)-(c) Estimate of the specific heat anomaly in FeSe at the superconducting transition, $\Delta C/T$, in the pressure regimes 0\,GPa$\,\le\,p\,\le\,$0.84\,GPa ($p\,<\,p_1$, a),  0.91\,GPa$\,\le\,p\,\le\,$1.58\,GPa ($p_1\,<\,p\,<\,p_2$, b) and  1.72\,GPa$\,\le\,p\,\le\,$2.36\,GPa ($p\,>\,p_2$, c). The inset of (c) shows a blow-up of the data set in the main panel. The dotted lines in the inset of (a) indicate exemplarily the equal-area construction in $\Delta C/T$ used to determine the superconducting jump size $\Delta C_{sc}/T_c$ and  the critical temperature $T_c$; (d) Evolution of $\Delta C_{sc}/T_c$ (left axis) as well as superconducting transition width (right axis; see Fig.\,S10) as a function of $p$. Purple bars indicate the position of critical pressures $p_1$ and $p_2$; (e) $\Delta C_{sc}/T_c$ as a function of the ratio $T_c/T_M$. Black circles (grey triangles) correspond to data in the pressure regime $p_1\,<\,p\,<\,p_2$ ($p\,>\,p_2$).}
	\label{fig:scjump}
	\end{figure}		
	\end{center}
	
	Next, we present in Fig.\,\ref{fig:scjump} the evolution of the specific heat jump across the superconducting transition at $T_c$ in the three distinct pressure regimes (a) $p\,<\,p_1$, (b) $p_1\,<\,p\,<\,p_2$ and (c) $p\,>\,p_2$ (see Figs.\,S8 and S9 for raw data). At all $p$ up to 2.36\,GPa, we resolve a clear specific heat anomaly at low $T$, associated with the superconducting transition at $T_c$. To determine $T_c$ and the superconducting jump size $\Delta C_{sc}/T_c(p)$, we use an equal-area construction in $\Delta C/T$ (see dotted lines in inset of Fig.\,\ref{fig:scjump}\,(a)). For $p\,\lesssim\,p_1$, we find an increase of $T_c$ together with an increase of $\Delta C_{sc}/T_c$ (see Fig.\,\ref{fig:scjump}\,(a)). Soon after the onset of magnetism at $p_1$, $T_c$ and $\Delta C_{sc}/T_c$ are suppressed with $p$ for $p\,<\,p_2$. Above $p_2$, $T_c$ increases slowly, however, $\Delta C_{sc}/T_c$ continues to be monotonically suppressed with increasing $p$. 
	
	Remarkably, we also find a sudden change of the shape of the $\Delta C/T(T_c)$ anomaly from almost mean-field-like at $p\,<\,p_1$ to a more $\lambda$-like shape with an extended high-$T$ tail at $p\,>\,p_1$. This change can be quantified in terms of a broadening parameter (see Fig.\,S10) which defines the width of superconducting transition and is shown in Fig.\,\ref{fig:scjump}\,(d) (right axis): it is almost constant as a function of $p$ for $p\,<\,p_1$, then exhibits a clear jump at $p_1$ (see also Fig.\,S11) and levels off again, until it increases rapidly for $p\,>\,p_2$. We stress that such sudden changes in the broadening, as observed here at $p_1$ and \textit{again} at $p_2$, are unlikely to result from pressure inhomogeneities related to the freezing of the pressure medium \cite{Torikachvili15}, and therefore rather reflect a change of intrinsic physics of FeSe.
	
\begin{center}
	\begin{figure}[h!]
	\includegraphics[width=\columnwidth]{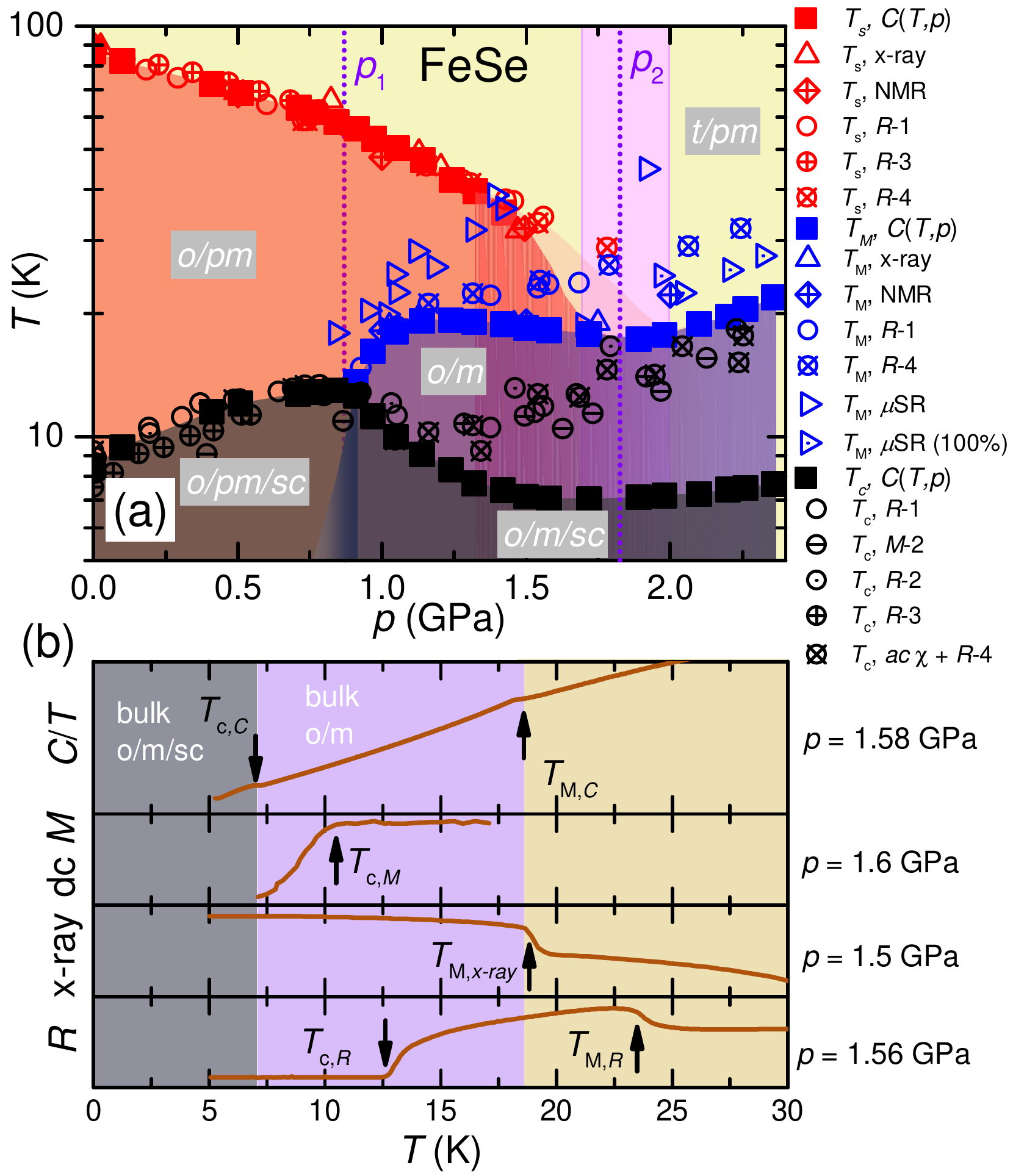} 
	\caption{(a) Temperature-pressure phase diagram of FeSe, determined from specific heat measurements $C(T,p)$ (full squares). Red symbols correspond to the structural transition temperature $T_s$, black symbols to the superconducting transition temperature $T_c$ and blue symbols to the magnetic transition temperature $T_M$. The phase regions are labeled by \textit{t/pm} (tetragonal/paramagnetic; light yellow), \textit{o/pm} (orthorhombic/paramagnetic; red), \textit{o/m} (orthorhombic/magnetic; blue) and \textit{sc} (superconducting; brown/grey). Purple dotted vertical lines mark two characteristic pressures, $p_1$ and $p_2$. The error in the determination of $p_2$ is indicated by the light purple bar. The specific heat data is contrasted with data from various other techniques from literature, i.e., x-ray scattering \cite{Kothapalli16}, NMR\cite{Wiecki17}, resistance and magnetization ($R$-1\cite{Kaluarachchi16}, $R$-2\cite{Miyoshi14} and $M$-2\cite{Miyoshi14}, $R$-3\cite{Knoener15}, $R$-4\cite{Terashima15}), and $\mu$SR \cite{Bendele12,Khasanov18}; (b) Comparison of $C/T$ data at 1.58 GPa to $M$\cite{Miyoshi14} data, x-ray data of the orthorhombic distortion \cite{Kothapalli16} and $R$\cite{Kaluarachchi16} ($R$) data at similar nominal pressures.}
	\label{fig:phasediagram}
	\end{figure}		
	\end{center}

	
	We can now proceed with discussing the two central results of this study. The first one relates to the question of bulk superconductivity in FeSe under $p$ and its relationship with magnetism. Here, the observation of a finite $\Delta C_{sc}/T_c$ at all $p$ speaks in strong favor of bulk superconductivity in FeSe, which coexists with nematic order at low $p$ as well as with magnetic-nematic order at high $p$. The fact that $\Delta C_{sc}/T_c$, which, in simple BCS theory, is a measure of the superconducting condensation energy, is strongly suppressed with $p$ for $p\,\gtrsim\,p_1$ (see Fig.\,\ref{fig:scjump}\,(d)) indicates that magnetism competes with superconductivity in FeSe, resulting in either microscopic coexistence  or in a macroscopic phase segregation \cite{Cheung18}. Importantly, competition is also the case for the region $p\,>\,p_2$, even though $T_c$ and $T_M$ both increase with $p$. This unusual possibility is included in an earlier model \cite{Machida81} on competing spin-density wave and superconducting order in itinerant systems, which provides the general tendency that competition leads to a decrease of $T_c/T_M$ (rather than a decrease of $T_c$ itself), when $T_M$ is increased. Our specific heat results of the bulk $T_M$ and $T_c$ values (see Figs.\,\ref{fig:phasediagram} and S1\,(a)) indeed show that this is the case in FeSe at high $p$: notably, $\Delta C_{sc}/T_c$ is suppressed with decreasing $T_c/T_M$ (see Fig.\,\ref{fig:scjump}\,(e)). Therefore, our results strengthen the similarities of FeSe to other Fe-based superconductors \cite{Canfield10,Rotter09,Pratt09,Christianson09,Luo12,Fernandes10,Boehmer15b,Budko18,Cheung18}.
	
	The second result is summarized in the $T$-$p$-phase diagram in Fig.\,\ref{fig:phasediagram} (a) (see Fig.\,S1 for simplified versions of this phase diagram). In this figure, we compare the transition temperatures $T_s$, $T_M$ and $T_c$ from the present $C(T,p)$ work (full symbols), with those reported in literature\cite{footnote}, based on x-ray scattering \cite{Kothapalli16,Boehmer18}, NMR \cite{Wiecki17}, resistance \cite{Kaluarachchi16,Miyoshi14,Knoener15,Terashima15}, magnetization \cite{Terashima15} and $\mu$SR \cite{Bendele12,Khasanov18} (open symbols). Surprisingly, whereas the majority of $T_s$ values and $T_c$ values for $p\,<\,p_1$, as well as the $p_1$ values themselves, are rather consistent, the $T_M$ and $T_c$ values for $p\,>\,p_1$ show strong discrepancies. Given that specific heat measurements provide the bulk, thermodynamic (and static) transition temperatures, we suggest below one possible way to rationalize these findings is in terms of superconducting and magnetic fluctuations which exist for $p\,\geq\,p_1$ over a wide $T$ range above $T_c$ and $T_M$, respectively.
	
	In terms of superconductivity for $p\,>\,p_1$, not only is the discrepancy of bulk $T_c$ values from the present study ($T_{c,C}$) and those from previous reports from transport and susceptibility ($T_{c,R/\chi}\,\gg\,T_{c,C}$, Fig.\,\ref{fig:phasediagram} (a) and (b)) remarkable, but it must be recalled that there is a simultaneous, sudden change in the shape of the $C$ anomaly at $p_1$, depicted in Fig.\,\ref{fig:scjump}. A sudden increase in broadening of the feature at $T_c$ at $p_1$ was also observed in other quantities \cite{Terashima15,Chen19}, such as resistance, despite being much larger there. Contrary to changes in transport features, though, the observed change in the specific heat feature is considered as a well-established signature \cite{Junod99,Adachi17,Yang17} of superconducting fluctuations \cite{Kasahara16} above the mean-field $T_c$. In this situation, the onset of diamagnetism \cite{Adachi17,Watashige17} at $T_{c,\chi}$ is likely found at higher temperatures than the bulk $T_{c,C}$, consistent with our results. Revisiting susceptibility data\cite{Miyoshi14,Terashima15} demonstrates that the bulk $T_{c,C}$ actually corresponds to the temperature at which FeSe exhibits saturating diamagnetism (see Fig.\,\ref{fig:phasediagram} (b)). Thus, a comparison of onset $T_{c,\chi}$ and $T_{c,C}$ can be used to estimate the $T$ range in which superconducting fluctuations exist. This $T$ range is small, but present for $p_1\,<\,p\,<\,p_2$ and it increases rapidly above $p_2$ ($\approx\,$10\,K$\,\simeq\,2 T_c$ at 2.36\,GPa, see Figs.\,\ref{fig:phasediagram}, S1 and S13). This mirrors the observed broadening of the $\Delta C/T$ feature at $T_c$.  Taken together, all these observations are consistent with a picture, in which significant changes of the Fermi surface \cite{Kaluarachchi16,Terashima16,Massat18} at $p_1$  and $p_2$ increase the $T$-range of fluctuations. Such extended fluctuations in the presence of competing magnetic order, suggested in the present work, might also naturally account for the absence of pronounced features at $T_c$ in microscopic NMR data \cite{Wang16} at $p\,>\,p_2$.

Concerning the magnetic transition, we find that the $T_M$ values from $C(T,p)$ are at the lower bound of values reported so far. It is remarkable, though, that similar $T_M$ values were inferred using the same technique in different studies (see, e.g., the two sets of open blue circles from resistance studies in Fig.\,\ref{fig:phasediagram}). This argues against experimental artifacts arising from a combination of different samples with slightly different stochiometry and different pressure media being solely responsible for the discrepancy in $T_M$ values. Instead, it seems likely that the observed spread in $T_M$ is related to the time scale of each experiment, ranging from $\sim\,\mu s$ for $\mu$SR \cite{Bendele12,Khasanov18} up to $\sim\,s$ for NMR \cite{Wang16,Wiecki17} up to static for $C(T)$ and x-ray probes (measuring the increase of orthorhombicity associated with the development of long-range order \cite{Kothapalli16}). We refrain from including the $T_M$ values inferred from the resistance in the present discussion, as the associated time scale, given by the scattering time, cannot be unequivocally defined. As $T_M(p)$ from the two static probes ($C(T)$ and x-ray) fall on top of each other ($T_{M,C}\,\simeq\,T_{M,x-ray}$, see Fig.\,\ref{fig:phasediagram}\,(b) and S1\,(b)) and $T_{M,C}\,\lesssim\,T_{M,NMR}\,\lesssim\,T_{\mu SR}$ at any given $p$, this is highly suggestive of magnetic fluctuations existing far above the static $T_{M,C}$. The extent in $T$ of these fluctuations above $T_M$ can be estimated from the spread of transition temperatures in Fig.\,\ref{fig:phasediagram}. This spread increases upon increasing $p$, even more rapidly above $p_2$, and reaches more than $\,\approx\,30\,$K above 2 GPa. The width of the specific heat peak at $T_M$ (see Fig.\,\ref{fig:data-magnetic-structural} (c)) provides further support for this statement, as it shows a progressive increase above $p_2$ (see Fig.\,S12), which reflects a sizable loss of magnetic entropy preceding the bulk $T_{M,C}$ upon cooling.

	Another scenario which could give rise to a similar phenomenology of the $T$-$p$ phase diagram, as well as the specific heat features, invokes electronic inhomogeneity \cite{Martiny19} giving rise to a spatially-fluctuating state. It is important to note though, that this inhomogeneity then must be intrinsically induced by the occurrence of magnetism, as evident from our phase diagram in Fig.\,\ref{fig:phasediagram}. It could, e.g., arise from the formation of domains in the magnetically-ordered state which are pinned by extrinsic disorder, inevitable in any real crystal. Whereas such a scenario certainly promotes a non-bulk superconducting state above $T_c$, causing zero resistance well above the bulk $T_c$ (such as the recently proposed fragile superconducting state \cite{Yu19}), it unlikely explains the correlation of time scales and transition temperatures for the magnetic transition. Thus, whereas for the superconducting transition either temporal or spatial fluctuations are consistent with our data, the results speak in favor of a temporal nature of the magnetic fluctuations.
	
	To verify which of these two scenarios is applicable in FeSe, it will be of crucial importance to identify the characteristic energy scales of the different orders in FeSe under pressure. One important key question here will be to resolve the magnetic structure of FeSe for $p\,>\,p_1$ which has still not been unequivocally determined to date. Nevertheless, we want to stress that our picture of the $T$-$p$ phase diagram of FeSe presents close similarity to the ones of the high-$T_c$ cuprate superconductors \cite{Keimer15}. In the latter case, there is growing evidence for the coexistence of superconductivity in the underdoped regime with other competing phases, such as charge-density waves \cite{SilvaNeto14} enhancing fluctuations \cite{Kivelson03,Torchinsky13} associated with both orders over a wide $T$ range above the respective bulk transition temperatures \cite{Keimer15,Vishik18}. Whereas this comparison is purely phenomenological at present, FeSe might serve as an important reference system to investigate the origin of such extended fluctuating regimes in the presence of competing orders, as superconductivity can be tuned through non-magnetic and magnetic states solely via pressure which does not introduce any additional disorder.

In conclusion, the presented specific heat data demonstrate that superconductivity is bulk in FeSe up to 2.36\,GPa, and competes with magnetism, whenever present. In the presence of magnetism, our results strongly suggest that superconducting and magnetic fluctuations exist over a wide temperature range above the respective bulk transition temperatures. This puts the phase diagram of FeSe under pressure in close similarity to those of underdoped cuprates in which the enhancement of phase fluctuations due to competing orders is considered as a key ingredient for high-$T_c$ superconductivity. 

\begin{acknowledgments}
We thank A. Kreyssig (Krey\ss ig), V. G. Kogan, D. Ryan and B. Andersen for useful discussions. In addition, we thank G. Drachuck for useful discussions and technical support with the ac specific heat setup in the initial stages of this work. Work at the Ames Laboratory was supported by the U.S. Department of Energy, Office of Science, Basic Energy Sciences, Materials Sciences and Engineering Division. The Ames Laboratory is operated for the U.S. Department of Energy by Iowa State University under Contract No. DEAC02-07CH11358.
\end{acknowledgments}

\bibliographystyle{apsrev}

\begin{thebibliography}
\expandafter\ifx\csname natexlab\endcsname\relax\def\natexlab#1{#1}\fi
\expandafter\ifx\csname bibnamefont\endcsname\relax
  \def\bibnamefont#1{#1}\fi
\expandafter\ifx\csname bibfnamefont\endcsname\relax
  \def\bibfnamefont#1{#1}\fi
\expandafter\ifx\csname citenamefont\endcsname\relax
  \def\citenamefont#1{#1}\fi
\expandafter\ifx\csname url\endcsname\relax
  \def\url#1{\texttt{#1}}\fi
\expandafter\ifx\csname urlprefix\endcsname\relax\def\urlprefix{URL }\fi
\providecommand{\bibinfo}[2]{#2}
\providecommand{\eprint}[2][]{\url{#2}}

\bibitem[{\citenamefont{Böhmer and Kreisel}(2017)}]{Boehmer17}
\bibinfo{author}{\bibfnamefont{A.~E.} \bibnamefont{Böhmer}} \bibnamefont{and}
  \bibinfo{author}{\bibfnamefont{A.}~\bibnamefont{Kreisel}},
  \bibinfo{journal}{Journal of Physics: Condensed Matter}
  \textbf{\bibinfo{volume}{30}}, \bibinfo{pages}{023001}
  (\bibinfo{year}{2017}).

\bibitem[{\citenamefont{Coldea and Watson}(2018)}]{Coldea18}
\bibinfo{author}{\bibfnamefont{A.~I.} \bibnamefont{Coldea}} \bibnamefont{and}
  \bibinfo{author}{\bibfnamefont{M.~D.} \bibnamefont{Watson}},
  \bibinfo{journal}{Annual Review of Condensed Matter Physics}
  \textbf{\bibinfo{volume}{9}}, \bibinfo{pages}{125} (\bibinfo{year}{2018}).

\bibitem[{\citenamefont{Paglione and Greene}(2010)}]{Paglione10}
\bibinfo{author}{\bibfnamefont{J.}~\bibnamefont{Paglione}} \bibnamefont{and}
  \bibinfo{author}{\bibfnamefont{R.~L.} \bibnamefont{Greene}},
  \bibinfo{journal}{Nature Physics} \textbf{\bibinfo{volume}{6}},
  \bibinfo{pages}{645–658} (\bibinfo{year}{2010}).

\bibitem[{\citenamefont{Johnston}(2010)}]{Johnston10}
\bibinfo{author}{\bibfnamefont{D.~C.} \bibnamefont{Johnston}},
  \bibinfo{journal}{Advances in Physics} \textbf{\bibinfo{volume}{59}},
  \bibinfo{pages}{803} (\bibinfo{year}{2010}).

\bibitem[{\citenamefont{Stewart}(2011)}]{Stewart11}
\bibinfo{author}{\bibfnamefont{G.~R.} \bibnamefont{Stewart}},
  \bibinfo{journal}{Rev. Mod. Phys.} \textbf{\bibinfo{volume}{83}},
  \bibinfo{pages}{1589} (\bibinfo{year}{2011}).

\bibitem[{\citenamefont{Hosono and Kuroki}(2015)}]{Hosono15}
\bibinfo{author}{\bibfnamefont{H.}~\bibnamefont{Hosono}} \bibnamefont{and}
  \bibinfo{author}{\bibfnamefont{K.}~\bibnamefont{Kuroki}},
  \bibinfo{journal}{Physica C: Superconductivity and its Applications}
  \textbf{\bibinfo{volume}{514}}, \bibinfo{pages}{399 } (\bibinfo{year}{2015}).

\bibitem[{\citenamefont{Canfield and Bud'ko}(2010)}]{Canfield10}
\bibinfo{author}{\bibfnamefont{P.~C.} \bibnamefont{Canfield}} \bibnamefont{and}
  \bibinfo{author}{\bibfnamefont{S.~L.} \bibnamefont{Bud'ko}},
  \bibinfo{journal}{Annual Review of Condensed Matter Physics}
  \textbf{\bibinfo{volume}{1}}, \bibinfo{pages}{27} (\bibinfo{year}{2010}).

\bibitem[{\citenamefont{Hsu et~al.}(2008)\citenamefont{Hsu, Luo, Yeh, Chen,
  Huang, Wu, Lee, Huang, Chu, Yan et~al.}}]{Hsu08}
\bibinfo{author}{\bibfnamefont{F.-C.} \bibnamefont{Hsu}},
  \bibinfo{author}{\bibfnamefont{J.-Y.} \bibnamefont{Luo}},
  \bibinfo{author}{\bibfnamefont{K.-W.} \bibnamefont{Yeh}},
  \bibinfo{author}{\bibfnamefont{T.-K.} \bibnamefont{Chen}},
  \bibinfo{author}{\bibfnamefont{T.-W.} \bibnamefont{Huang}},
  \bibinfo{author}{\bibfnamefont{P.~M.} \bibnamefont{Wu}},
  \bibinfo{author}{\bibfnamefont{Y.-C.} \bibnamefont{Lee}},
  \bibinfo{author}{\bibfnamefont{Y.-L.} \bibnamefont{Huang}},
  \bibinfo{author}{\bibfnamefont{Y.-Y.} \bibnamefont{Chu}},
  \bibinfo{author}{\bibfnamefont{D.-C.} \bibnamefont{Yan}},
  \bibnamefont{et~al.}, \bibinfo{journal}{Proceedings of the National Academy
  of Sciences} \textbf{\bibinfo{volume}{105}}, \bibinfo{pages}{14262}
  (\bibinfo{year}{2008}).

\bibitem[{\citenamefont{Ge et~al.}(2015)\citenamefont{Ge, Liu, Liu, Gao, Qian,
  Xue, Liu, and Jia}}]{Ge15}
\bibinfo{author}{\bibfnamefont{J.-F.} \bibnamefont{Ge}},
  \bibinfo{author}{\bibfnamefont{Z.-L.} \bibnamefont{Liu}},
  \bibinfo{author}{\bibfnamefont{C.}~\bibnamefont{Liu}},
  \bibinfo{author}{\bibfnamefont{C.-L.} \bibnamefont{Gao}},
  \bibinfo{author}{\bibfnamefont{D.}~\bibnamefont{Qian}},
  \bibinfo{author}{\bibfnamefont{Q.-K.} \bibnamefont{Xue}},
  \bibinfo{author}{\bibfnamefont{Y.}~\bibnamefont{Liu}}, \bibnamefont{and}
  \bibinfo{author}{\bibfnamefont{J.-F.} \bibnamefont{Jia}},
  \bibinfo{journal}{Nature Materials} \textbf{\bibinfo{volume}{14}},
  \bibinfo{pages}{285–289} (\bibinfo{year}{2015}).

\bibitem[{\citenamefont{Sadovskii}(2016)}]{Sadovskii16}
\bibinfo{author}{\bibfnamefont{M.~V.} \bibnamefont{Sadovskii}},
  \bibinfo{journal}{Physics-Uspekhi} \textbf{\bibinfo{volume}{59}},
  \bibinfo{pages}{947} (\bibinfo{year}{2016}).

\bibitem[{\citenamefont{Wang et~al.}(2017)\citenamefont{Wang, Liu, Liu, and
  Wang}}]{Wang17}
\bibinfo{author}{\bibfnamefont{Z.}~\bibnamefont{Wang}},
  \bibinfo{author}{\bibfnamefont{C.}~\bibnamefont{Liu}},
  \bibinfo{author}{\bibfnamefont{Y.}~\bibnamefont{Liu}}, \bibnamefont{and}
  \bibinfo{author}{\bibfnamefont{J.}~\bibnamefont{Wang}},
  \bibinfo{journal}{Journal of Physics: Condensed Matter}
  \textbf{\bibinfo{volume}{29}}, \bibinfo{pages}{153001}
  (\bibinfo{year}{2017}).

\bibitem[{\citenamefont{Huang and Hoffman}(2017)}]{Huang17}
\bibinfo{author}{\bibfnamefont{D.}~\bibnamefont{Huang}} \bibnamefont{and}
  \bibinfo{author}{\bibfnamefont{J.~E.} \bibnamefont{Hoffman}},
  \bibinfo{journal}{Annual Review of Condensed Matter Physics}
  \textbf{\bibinfo{volume}{8}}, \bibinfo{pages}{311} (\bibinfo{year}{2017}).

\bibitem[{\citenamefont{Burrard-Lucas et~al.}(2013)\citenamefont{Burrard-Lucas,
  Free, Sedlmaier, Wright, Cassidy, Hara, Corkett, Lancaster, Baker, Blundell
  et~al.}}]{Burrard-Lucas13}
\bibinfo{author}{\bibfnamefont{M.}~\bibnamefont{Burrard-Lucas}},
  \bibinfo{author}{\bibfnamefont{D.~G.} \bibnamefont{Free}},
  \bibinfo{author}{\bibfnamefont{S.~J.} \bibnamefont{Sedlmaier}},
  \bibinfo{author}{\bibfnamefont{J.~D.} \bibnamefont{Wright}},
  \bibinfo{author}{\bibfnamefont{S.~J.} \bibnamefont{Cassidy}},
  \bibinfo{author}{\bibfnamefont{Y.}~\bibnamefont{Hara}},
  \bibinfo{author}{\bibfnamefont{A.~J.} \bibnamefont{Corkett}},
  \bibinfo{author}{\bibfnamefont{T.}~\bibnamefont{Lancaster}},
  \bibinfo{author}{\bibfnamefont{P.~J.} \bibnamefont{Baker}},
  \bibinfo{author}{\bibfnamefont{S.~J.} \bibnamefont{Blundell}},
  \bibnamefont{et~al.}, \bibinfo{journal}{Nature Materials}
  \textbf{\bibinfo{volume}{12}}, \bibinfo{pages}{15–19}
  (\bibinfo{year}{2013}).

\bibitem[{\citenamefont{Mizuguchi et~al.}(2008)\citenamefont{Mizuguchi,
  Tomioka, Tsuda, Yamaguchi, and Takano}}]{Mizuguchi08}
\bibinfo{author}{\bibfnamefont{Y.}~\bibnamefont{Mizuguchi}},
  \bibinfo{author}{\bibfnamefont{F.}~\bibnamefont{Tomioka}},
  \bibinfo{author}{\bibfnamefont{S.}~\bibnamefont{Tsuda}},
  \bibinfo{author}{\bibfnamefont{T.}~\bibnamefont{Yamaguchi}},
  \bibnamefont{and} \bibinfo{author}{\bibfnamefont{Y.}~\bibnamefont{Takano}},
  \bibinfo{journal}{Applied Physics Letters} \textbf{\bibinfo{volume}{93}},
  \bibinfo{pages}{152505} (\bibinfo{year}{2008}).

\bibitem[{\citenamefont{Medvedev et~al.}(2090)\citenamefont{Medvedev, McQueen,
  Troyan, Palasyuk, Eremets, Cava, Naghavi, Casper, Ksenofontov, Wortmann
  et~al.}}]{Medvedev09}
\bibinfo{author}{\bibfnamefont{S.}~\bibnamefont{Medvedev}},
  \bibinfo{author}{\bibfnamefont{T.~M.} \bibnamefont{McQueen}},
  \bibinfo{author}{\bibfnamefont{I.~A.} \bibnamefont{Troyan}},
  \bibinfo{author}{\bibfnamefont{T.}~\bibnamefont{Palasyuk}},
  \bibinfo{author}{\bibfnamefont{M.~I.} \bibnamefont{Eremets}},
  \bibinfo{author}{\bibfnamefont{R.~J.} \bibnamefont{Cava}},
  \bibinfo{author}{\bibfnamefont{S.}~\bibnamefont{Naghavi}},
  \bibinfo{author}{\bibfnamefont{F.}~\bibnamefont{Casper}},
  \bibinfo{author}{\bibfnamefont{V.}~\bibnamefont{Ksenofontov}},
  \bibinfo{author}{\bibfnamefont{G.}~\bibnamefont{Wortmann}},
  \bibnamefont{et~al.}, \bibinfo{journal}{Nature Materials}
  \textbf{\bibinfo{volume}{8}}, \bibinfo{pages}{630–633}
  (\bibinfo{year}{2090}).

\bibitem[{\citenamefont{Margadonna et~al.}(2009)\citenamefont{Margadonna,
  Takabayashi, Ohishi, Mizuguchi, Takano, Kagayama, Nakagawa, Takata, and
  Prassides}}]{Margadonna09}
\bibinfo{author}{\bibfnamefont{S.}~\bibnamefont{Margadonna}},
  \bibinfo{author}{\bibfnamefont{Y.}~\bibnamefont{Takabayashi}},
  \bibinfo{author}{\bibfnamefont{Y.}~\bibnamefont{Ohishi}},
  \bibinfo{author}{\bibfnamefont{Y.}~\bibnamefont{Mizuguchi}},
  \bibinfo{author}{\bibfnamefont{Y.}~\bibnamefont{Takano}},
  \bibinfo{author}{\bibfnamefont{T.}~\bibnamefont{Kagayama}},
  \bibinfo{author}{\bibfnamefont{T.}~\bibnamefont{Nakagawa}},
  \bibinfo{author}{\bibfnamefont{M.}~\bibnamefont{Takata}}, \bibnamefont{and}
  \bibinfo{author}{\bibfnamefont{K.}~\bibnamefont{Prassides}},
  \bibinfo{journal}{Phys. Rev. B} \textbf{\bibinfo{volume}{80}},
  \bibinfo{pages}{064506} (\bibinfo{year}{2009}).

\bibitem[{\citenamefont{Garbarino et~al.}(2009)\citenamefont{Garbarino, Sow,
  Lejay, Sulpice, Toulemonde, Mezouar, and
  N{\'{u}}{\~{n}}ez-Regueiro}}]{Garbarino09}
\bibinfo{author}{\bibfnamefont{G.}~\bibnamefont{Garbarino}},
  \bibinfo{author}{\bibfnamefont{A.}~\bibnamefont{Sow}},
  \bibinfo{author}{\bibfnamefont{P.}~\bibnamefont{Lejay}},
  \bibinfo{author}{\bibfnamefont{A.}~\bibnamefont{Sulpice}},
  \bibinfo{author}{\bibfnamefont{P.}~\bibnamefont{Toulemonde}},
  \bibinfo{author}{\bibfnamefont{M.}~\bibnamefont{Mezouar}}, \bibnamefont{and}
  \bibinfo{author}{\bibfnamefont{M.}~\bibnamefont{N{\'{u}}{\~{n}}ez-Regueiro}},
  \bibinfo{journal}{{EPL} (Europhysics Letters)} \textbf{\bibinfo{volume}{86}},
  \bibinfo{pages}{27001} (\bibinfo{year}{2009}).

\bibitem[{\citenamefont{Masaki et~al.}(2009)\citenamefont{Masaki, Kotegawa,
  Hara, Tou, Murata, Mizuguchi, and Takano}}]{Masaki09}
\bibinfo{author}{\bibfnamefont{S.}~\bibnamefont{Masaki}},
  \bibinfo{author}{\bibfnamefont{H.}~\bibnamefont{Kotegawa}},
  \bibinfo{author}{\bibfnamefont{Y.}~\bibnamefont{Hara}},
  \bibinfo{author}{\bibfnamefont{H.}~\bibnamefont{Tou}},
  \bibinfo{author}{\bibfnamefont{K.}~\bibnamefont{Murata}},
  \bibinfo{author}{\bibfnamefont{Y.}~\bibnamefont{Mizuguchi}},
  \bibnamefont{and} \bibinfo{author}{\bibfnamefont{Y.}~\bibnamefont{Takano}},
  \bibinfo{journal}{Journal of the Physical Society of Japan}
  \textbf{\bibinfo{volume}{78}}, \bibinfo{pages}{063704}
  (\bibinfo{year}{2009}).

\bibitem[{\citenamefont{Okabe et~al.}(2010)\citenamefont{Okabe, Takeshita,
  Horigane, Muranaka, and Akimitsu}}]{Okabe10}
\bibinfo{author}{\bibfnamefont{H.}~\bibnamefont{Okabe}},
  \bibinfo{author}{\bibfnamefont{N.}~\bibnamefont{Takeshita}},
  \bibinfo{author}{\bibfnamefont{K.}~\bibnamefont{Horigane}},
  \bibinfo{author}{\bibfnamefont{T.}~\bibnamefont{Muranaka}}, \bibnamefont{and}
  \bibinfo{author}{\bibfnamefont{J.}~\bibnamefont{Akimitsu}},
  \bibinfo{journal}{Phys. Rev. B} \textbf{\bibinfo{volume}{81}},
  \bibinfo{pages}{205119} (\bibinfo{year}{2010}).

\bibitem[{\citenamefont{Margadonna et~al.}(2008)\citenamefont{Margadonna,
  Takabayashi, McDonald, Kasperkiewicz, Mizuguchi, Takano, Fitch, Suard, and
  Prassides}}]{Margadonna08}
\bibinfo{author}{\bibfnamefont{S.}~\bibnamefont{Margadonna}},
  \bibinfo{author}{\bibfnamefont{Y.}~\bibnamefont{Takabayashi}},
  \bibinfo{author}{\bibfnamefont{M.~T.} \bibnamefont{McDonald}},
  \bibinfo{author}{\bibfnamefont{K.}~\bibnamefont{Kasperkiewicz}},
  \bibinfo{author}{\bibfnamefont{Y.}~\bibnamefont{Mizuguchi}},
  \bibinfo{author}{\bibfnamefont{Y.}~\bibnamefont{Takano}},
  \bibinfo{author}{\bibfnamefont{A.~N.} \bibnamefont{Fitch}},
  \bibinfo{author}{\bibfnamefont{E.}~\bibnamefont{Suard}}, \bibnamefont{and}
  \bibinfo{author}{\bibfnamefont{K.}~\bibnamefont{Prassides}},
  \bibinfo{journal}{Chem. Commun.} pp. \bibinfo{pages}{5607--5609}
  (\bibinfo{year}{2008}).

\bibitem[{\citenamefont{McQueen et~al.}(2009)\citenamefont{McQueen, Williams,
  Stephens, Tao, Zhu, Ksenofontov, Casper, Felser, and Cava}}]{McQueen09}
\bibinfo{author}{\bibfnamefont{T.~M.} \bibnamefont{McQueen}},
  \bibinfo{author}{\bibfnamefont{A.~J.} \bibnamefont{Williams}},
  \bibinfo{author}{\bibfnamefont{P.~W.} \bibnamefont{Stephens}},
  \bibinfo{author}{\bibfnamefont{J.}~\bibnamefont{Tao}},
  \bibinfo{author}{\bibfnamefont{Y.}~\bibnamefont{Zhu}},
  \bibinfo{author}{\bibfnamefont{V.}~\bibnamefont{Ksenofontov}},
  \bibinfo{author}{\bibfnamefont{F.}~\bibnamefont{Casper}},
  \bibinfo{author}{\bibfnamefont{C.}~\bibnamefont{Felser}}, \bibnamefont{and}
  \bibinfo{author}{\bibfnamefont{R.~J.} \bibnamefont{Cava}},
  \bibinfo{journal}{Phys. Rev. Lett.} \textbf{\bibinfo{volume}{103}},
  \bibinfo{pages}{057002} (\bibinfo{year}{2009}).

\bibitem[{\citenamefont{B\"ohmer et~al.}(2015)\citenamefont{B\"ohmer, Arai,
  Hardy, Hattori, Iye, Wolf, L\"ohneysen, Ishida, and Meingast}}]{Boehmer15}
\bibinfo{author}{\bibfnamefont{A.~E.} \bibnamefont{B\"ohmer}},
  \bibinfo{author}{\bibfnamefont{T.}~\bibnamefont{Arai}},
  \bibinfo{author}{\bibfnamefont{F.}~\bibnamefont{Hardy}},
  \bibinfo{author}{\bibfnamefont{T.}~\bibnamefont{Hattori}},
  \bibinfo{author}{\bibfnamefont{T.}~\bibnamefont{Iye}},
  \bibinfo{author}{\bibfnamefont{T.}~\bibnamefont{Wolf}},
  \bibinfo{author}{\bibfnamefont{H.~v.} \bibnamefont{L\"ohneysen}},
  \bibinfo{author}{\bibfnamefont{K.}~\bibnamefont{Ishida}}, \bibnamefont{and}
  \bibinfo{author}{\bibfnamefont{C.}~\bibnamefont{Meingast}},
  \bibinfo{journal}{Phys. Rev. Lett.} \textbf{\bibinfo{volume}{114}},
  \bibinfo{pages}{027001} (\bibinfo{year}{2015}).

\bibitem[{\citenamefont{Watson et~al.}(2015)\citenamefont{Watson, Kim,
  Haghighirad, Davies, McCollam, Narayanan, Blake, Chen, Ghannadzadeh,
  Schofield et~al.}}]{Watson15}
\bibinfo{author}{\bibfnamefont{M.~D.} \bibnamefont{Watson}},
  \bibinfo{author}{\bibfnamefont{T.~K.} \bibnamefont{Kim}},
  \bibinfo{author}{\bibfnamefont{A.~A.} \bibnamefont{Haghighirad}},
  \bibinfo{author}{\bibfnamefont{N.~R.} \bibnamefont{Davies}},
  \bibinfo{author}{\bibfnamefont{A.}~\bibnamefont{McCollam}},
  \bibinfo{author}{\bibfnamefont{A.}~\bibnamefont{Narayanan}},
  \bibinfo{author}{\bibfnamefont{S.~F.} \bibnamefont{Blake}},
  \bibinfo{author}{\bibfnamefont{Y.~L.} \bibnamefont{Chen}},
  \bibinfo{author}{\bibfnamefont{S.}~\bibnamefont{Ghannadzadeh}},
  \bibinfo{author}{\bibfnamefont{A.~J.} \bibnamefont{Schofield}},
  \bibnamefont{et~al.}, \bibinfo{journal}{Phys. Rev. B}
  \textbf{\bibinfo{volume}{91}}, \bibinfo{pages}{155106}
  (\bibinfo{year}{2015}).

\bibitem[{\citenamefont{Tanatar et~al.}(2016)\citenamefont{Tanatar, B\"ohmer,
  Timmons, Sch\"utt, Drachuck, Taufour, Kothapalli, Kreyssig, Bud'ko, Canfield
  et~al.}}]{Tanatar16}
\bibinfo{author}{\bibfnamefont{M.~A.} \bibnamefont{Tanatar}},
  \bibinfo{author}{\bibfnamefont{A.~E.} \bibnamefont{B\"ohmer}},
  \bibinfo{author}{\bibfnamefont{E.~I.} \bibnamefont{Timmons}},
  \bibinfo{author}{\bibfnamefont{M.}~\bibnamefont{Sch\"utt}},
  \bibinfo{author}{\bibfnamefont{G.}~\bibnamefont{Drachuck}},
  \bibinfo{author}{\bibfnamefont{V.}~\bibnamefont{Taufour}},
  \bibinfo{author}{\bibfnamefont{K.}~\bibnamefont{Kothapalli}},
  \bibinfo{author}{\bibfnamefont{A.}~\bibnamefont{Kreyssig}},
  \bibinfo{author}{\bibfnamefont{S.~L.} \bibnamefont{Bud'ko}},
  \bibinfo{author}{\bibfnamefont{P.~C.} \bibnamefont{Canfield}},
  \bibnamefont{et~al.}, \bibinfo{journal}{Phys. Rev. Lett.}
  \textbf{\bibinfo{volume}{117}}, \bibinfo{pages}{127001}
  (\bibinfo{year}{2016}).

\bibitem[{\citenamefont{Baek et~al.}(2015)\citenamefont{Baek, Efremov, Ok, Kim,
  van~den Brink, and Büchner}}]{Baek15}
\bibinfo{author}{\bibfnamefont{S.-H.} \bibnamefont{Baek}},
  \bibinfo{author}{\bibfnamefont{D.~V.} \bibnamefont{Efremov}},
  \bibinfo{author}{\bibfnamefont{J.~M.} \bibnamefont{Ok}},
  \bibinfo{author}{\bibfnamefont{J.~S.} \bibnamefont{Kim}},
  \bibinfo{author}{\bibfnamefont{J.}~\bibnamefont{van~den Brink}},
  \bibnamefont{and} \bibinfo{author}{\bibfnamefont{B.}~\bibnamefont{Büchner}},
  \bibinfo{journal}{Nature Materials} \textbf{\bibinfo{volume}{14}},
  \bibinfo{pages}{210–214} (\bibinfo{year}{2015}).

\bibitem[{\citenamefont{Fernandes et~al.}(2014)\citenamefont{Fernandes,
  Chubukov, and Schmalian}}]{Fernandes14}
\bibinfo{author}{\bibfnamefont{R.~M.} \bibnamefont{Fernandes}},
  \bibinfo{author}{\bibfnamefont{A.~V.} \bibnamefont{Chubukov}},
  \bibnamefont{and}
  \bibinfo{author}{\bibfnamefont{J.}~\bibnamefont{Schmalian}},
  \bibinfo{journal}{Nature Physics} \textbf{\bibinfo{volume}{10}},
  \bibinfo{pages}{97–104} (\bibinfo{year}{2014}).

\bibitem[{\citenamefont{Bendele et~al.}(2010)\citenamefont{Bendele, Amato,
  Conder, Elender, Keller, Klauss, Luetkens, Pomjakushina, Raselli, and
  Khasanov}}]{Bendele10}
\bibinfo{author}{\bibfnamefont{M.}~\bibnamefont{Bendele}},
  \bibinfo{author}{\bibfnamefont{A.}~\bibnamefont{Amato}},
  \bibinfo{author}{\bibfnamefont{K.}~\bibnamefont{Conder}},
  \bibinfo{author}{\bibfnamefont{M.}~\bibnamefont{Elender}},
  \bibinfo{author}{\bibfnamefont{H.}~\bibnamefont{Keller}},
  \bibinfo{author}{\bibfnamefont{H.-H.} \bibnamefont{Klauss}},
  \bibinfo{author}{\bibfnamefont{H.}~\bibnamefont{Luetkens}},
  \bibinfo{author}{\bibfnamefont{E.}~\bibnamefont{Pomjakushina}},
  \bibinfo{author}{\bibfnamefont{A.}~\bibnamefont{Raselli}}, \bibnamefont{and}
  \bibinfo{author}{\bibfnamefont{R.}~\bibnamefont{Khasanov}},
  \bibinfo{journal}{Phys. Rev. Lett.} \textbf{\bibinfo{volume}{104}},
  \bibinfo{pages}{087003} (\bibinfo{year}{2010}).

\bibitem[{\citenamefont{Yi et~al.}(2015)\citenamefont{Yi, Liu, Zhang, Yu, Zhu,
  Lee, Moore, Schmitt, Li, Riggs et~al.}}]{Yi15}
\bibinfo{author}{\bibfnamefont{M.}~\bibnamefont{Yi}},
  \bibinfo{author}{\bibfnamefont{Z.-K.} \bibnamefont{Liu}},
  \bibinfo{author}{\bibfnamefont{Y.}~\bibnamefont{Zhang}},
  \bibinfo{author}{\bibfnamefont{R.}~\bibnamefont{Yu}},
  \bibinfo{author}{\bibfnamefont{J.-X.} \bibnamefont{Zhu}},
  \bibinfo{author}{\bibfnamefont{J.}~\bibnamefont{Lee}},
  \bibinfo{author}{\bibfnamefont{R.}~\bibnamefont{Moore}},
  \bibinfo{author}{\bibfnamefont{F.}~\bibnamefont{Schmitt}},
  \bibinfo{author}{\bibfnamefont{W.}~\bibnamefont{Li}},
  \bibinfo{author}{\bibfnamefont{S.}~\bibnamefont{Riggs}},
  \bibnamefont{et~al.}, \bibinfo{journal}{Nature Communications}
  \textbf{\bibinfo{volume}{6}}, \bibinfo{pages}{7777} (\bibinfo{year}{2015}).

\bibitem[{\citenamefont{Kasahara et~al.}(2014)\citenamefont{Kasahara,
  Watashige, Hanaguri, Kohsaka, Yamashita, Shimoyama, Mizukami, Endo, Ikeda,
  Aoyama et~al.}}]{Kasahara14}
\bibinfo{author}{\bibfnamefont{S.}~\bibnamefont{Kasahara}},
  \bibinfo{author}{\bibfnamefont{T.}~\bibnamefont{Watashige}},
  \bibinfo{author}{\bibfnamefont{T.}~\bibnamefont{Hanaguri}},
  \bibinfo{author}{\bibfnamefont{Y.}~\bibnamefont{Kohsaka}},
  \bibinfo{author}{\bibfnamefont{T.}~\bibnamefont{Yamashita}},
  \bibinfo{author}{\bibfnamefont{Y.}~\bibnamefont{Shimoyama}},
  \bibinfo{author}{\bibfnamefont{Y.}~\bibnamefont{Mizukami}},
  \bibinfo{author}{\bibfnamefont{R.}~\bibnamefont{Endo}},
  \bibinfo{author}{\bibfnamefont{H.}~\bibnamefont{Ikeda}},
  \bibinfo{author}{\bibfnamefont{K.}~\bibnamefont{Aoyama}},
  \bibnamefont{et~al.}, \bibinfo{journal}{Proceedings of the National Academy
  of Sciences} \textbf{\bibinfo{volume}{111}}, \bibinfo{pages}{16309}
  (\bibinfo{year}{2014}).

\bibitem[{\citenamefont{Kasahara et~al.}(2016)\citenamefont{Kasahara,
  Yamashita, Shi, Kobayashi, Shimoyama, Watashige, Ishida, Terashima, Wolf,
  Hardy et~al.}}]{Kasahara16}
\bibinfo{author}{\bibfnamefont{S.}~\bibnamefont{Kasahara}},
  \bibinfo{author}{\bibfnamefont{T.}~\bibnamefont{Yamashita}},
  \bibinfo{author}{\bibfnamefont{A.}~\bibnamefont{Shi}},
  \bibinfo{author}{\bibfnamefont{R.}~\bibnamefont{Kobayashi}},
  \bibinfo{author}{\bibfnamefont{Y.}~\bibnamefont{Shimoyama}},
  \bibinfo{author}{\bibfnamefont{T.}~\bibnamefont{Watashige}},
  \bibinfo{author}{\bibfnamefont{K.}~\bibnamefont{Ishida}},
  \bibinfo{author}{\bibfnamefont{T.}~\bibnamefont{Terashima}},
  \bibinfo{author}{\bibfnamefont{T.}~\bibnamefont{Wolf}},
  \bibinfo{author}{\bibfnamefont{F.}~\bibnamefont{Hardy}},
  \bibnamefont{et~al.}, \bibinfo{journal}{Nature Communications}
  \textbf{\bibinfo{volume}{7}}, \bibinfo{pages}{12843} (\bibinfo{year}{2016}).

\bibitem[{\citenamefont{Watashige et~al.}(2017)\citenamefont{Watashige,
  Arsenijević, Yamashita, Terazawa, Onishi, Opherden, Kasahara, Tokiwa,
  Kasahara, Shibauchi et~al.}}]{Watashige17}
\bibinfo{author}{\bibfnamefont{T.}~\bibnamefont{Watashige}},
  \bibinfo{author}{\bibfnamefont{S.}~\bibnamefont{Arsenijević}},
  \bibinfo{author}{\bibfnamefont{T.}~\bibnamefont{Yamashita}},
  \bibinfo{author}{\bibfnamefont{D.}~\bibnamefont{Terazawa}},
  \bibinfo{author}{\bibfnamefont{T.}~\bibnamefont{Onishi}},
  \bibinfo{author}{\bibfnamefont{L.}~\bibnamefont{Opherden}},
  \bibinfo{author}{\bibfnamefont{S.}~\bibnamefont{Kasahara}},
  \bibinfo{author}{\bibfnamefont{Y.}~\bibnamefont{Tokiwa}},
  \bibinfo{author}{\bibfnamefont{Y.}~\bibnamefont{Kasahara}},
  \bibinfo{author}{\bibfnamefont{T.}~\bibnamefont{Shibauchi}},
  \bibnamefont{et~al.}, \bibinfo{journal}{Journal of the Physical Society of
  Japan} \textbf{\bibinfo{volume}{86}}, \bibinfo{pages}{014707}
  (\bibinfo{year}{2017}).

\bibitem[{\citenamefont{Rinott et~al.}(2017)\citenamefont{Rinott, Chashka,
  Ribak, Rienks, Taleb-Ibrahimi, Fevre, Bertran, Randeria, and
  Kanigel1}}]{Rinott17}
\bibinfo{author}{\bibfnamefont{S.}~\bibnamefont{Rinott}},
  \bibinfo{author}{\bibfnamefont{K.~B.} \bibnamefont{Chashka}},
  \bibinfo{author}{\bibfnamefont{A.}~\bibnamefont{Ribak}},
  \bibinfo{author}{\bibfnamefont{E.~D.~L.} \bibnamefont{Rienks}},
  \bibinfo{author}{\bibfnamefont{A.}~\bibnamefont{Taleb-Ibrahimi}},
  \bibinfo{author}{\bibfnamefont{P.~L.} \bibnamefont{Fevre}},
  \bibinfo{author}{\bibfnamefont{F.}~\bibnamefont{Bertran}},
  \bibinfo{author}{\bibfnamefont{M.}~\bibnamefont{Randeria}}, \bibnamefont{and}
  \bibinfo{author}{\bibfnamefont{A.}~\bibnamefont{Kanigel1}},
  \bibinfo{journal}{Science Advances} \textbf{\bibinfo{volume}{3}},
  \bibinfo{pages}{e1602372} (\bibinfo{year}{2017}).

\bibitem[{\citenamefont{Hanaguri et~al.}(2019)\citenamefont{Hanaguri, Kasahara,
  B\"oker, Eremin, Shibauchi, and Matsuda}}]{Hanaguri19}
\bibinfo{author}{\bibfnamefont{T.}~\bibnamefont{Hanaguri}},
  \bibinfo{author}{\bibfnamefont{S.}~\bibnamefont{Kasahara}},
  \bibinfo{author}{\bibfnamefont{J.}~\bibnamefont{B\"oker}},
  \bibinfo{author}{\bibfnamefont{I.}~\bibnamefont{Eremin}},
  \bibinfo{author}{\bibfnamefont{T.}~\bibnamefont{Shibauchi}},
  \bibnamefont{and} \bibinfo{author}{\bibfnamefont{Y.}~\bibnamefont{Matsuda}},
  \bibinfo{journal}{Phys. Rev. Lett.} \textbf{\bibinfo{volume}{122}},
  \bibinfo{pages}{077001} (\bibinfo{year}{2019}).

\bibitem[{\citenamefont{Lubashevsky et~al.}(2012)\citenamefont{Lubashevsky,
  Lahoud, Chashka, Podolsky, and Kanigel}}]{Lubashevsky12}
\bibinfo{author}{\bibfnamefont{Y.}~\bibnamefont{Lubashevsky}},
  \bibinfo{author}{\bibfnamefont{E.}~\bibnamefont{Lahoud}},
  \bibinfo{author}{\bibfnamefont{K.}~\bibnamefont{Chashka}},
  \bibinfo{author}{\bibfnamefont{D.}~\bibnamefont{Podolsky}}, \bibnamefont{and}
  \bibinfo{author}{\bibfnamefont{A.}~\bibnamefont{Kanigel}},
  \bibinfo{journal}{Nature Physics} \textbf{\bibinfo{volume}{8}},
  \bibinfo{pages}{309–312} (\bibinfo{year}{2012}).

\bibitem[{\citenamefont{Kothapalli et~al.}(2016)\citenamefont{Kothapalli,
  Böhmer, Jayasekara, Ueland, Das, Sapkota, Taufour, Xiao, Alp, Bud’ko
  et~al.}}]{Kothapalli16}
\bibinfo{author}{\bibfnamefont{K.}~\bibnamefont{Kothapalli}},
  \bibinfo{author}{\bibfnamefont{A.~E.} \bibnamefont{Böhmer}},
  \bibinfo{author}{\bibfnamefont{W.~T.} \bibnamefont{Jayasekara}},
  \bibinfo{author}{\bibfnamefont{B.~G.} \bibnamefont{Ueland}},
  \bibinfo{author}{\bibfnamefont{P.}~\bibnamefont{Das}},
  \bibinfo{author}{\bibfnamefont{A.}~\bibnamefont{Sapkota}},
  \bibinfo{author}{\bibfnamefont{V.}~\bibnamefont{Taufour}},
  \bibinfo{author}{\bibfnamefont{Y.}~\bibnamefont{Xiao}},
  \bibinfo{author}{\bibfnamefont{E.}~\bibnamefont{Alp}},
  \bibinfo{author}{\bibfnamefont{S.~L.} \bibnamefont{Bud’ko}},
  \bibnamefont{et~al.}, \bibinfo{journal}{Nat. Commun.}
  \textbf{\bibinfo{volume}{7}}, \bibinfo{pages}{12728} (\bibinfo{year}{2016}).

\bibitem[{\citenamefont{Wang et~al.}(2016)\citenamefont{Wang, Sun, Cui, Song,
  Li, Yu, Lei, and Yu}}]{Wang16}
\bibinfo{author}{\bibfnamefont{P.~S.} \bibnamefont{Wang}},
  \bibinfo{author}{\bibfnamefont{S.~S.} \bibnamefont{Sun}},
  \bibinfo{author}{\bibfnamefont{Y.}~\bibnamefont{Cui}},
  \bibinfo{author}{\bibfnamefont{W.~H.} \bibnamefont{Song}},
  \bibinfo{author}{\bibfnamefont{T.~R.} \bibnamefont{Li}},
  \bibinfo{author}{\bibfnamefont{R.}~\bibnamefont{Yu}},
  \bibinfo{author}{\bibfnamefont{H.}~\bibnamefont{Lei}}, \bibnamefont{and}
  \bibinfo{author}{\bibfnamefont{W.}~\bibnamefont{Yu}}, \bibinfo{journal}{Phys.
  Rev. Lett.} \textbf{\bibinfo{volume}{117}}, \bibinfo{pages}{237001}
  (\bibinfo{year}{2016}).

\bibitem[{\citenamefont{Wiecki et~al.}(2017)\citenamefont{Wiecki, Nandi,
  B\"ohmer, Bud'ko, Canfield, and Furukawa}}]{Wiecki17}
\bibinfo{author}{\bibfnamefont{P.}~\bibnamefont{Wiecki}},
  \bibinfo{author}{\bibfnamefont{M.}~\bibnamefont{Nandi}},
  \bibinfo{author}{\bibfnamefont{A.~E.} \bibnamefont{B\"ohmer}},
  \bibinfo{author}{\bibfnamefont{S.~L.} \bibnamefont{Bud'ko}},
  \bibinfo{author}{\bibfnamefont{P.~C.} \bibnamefont{Canfield}},
  \bibnamefont{and} \bibinfo{author}{\bibfnamefont{Y.}~\bibnamefont{Furukawa}},
  \bibinfo{journal}{Phys. Rev. B} \textbf{\bibinfo{volume}{96}},
  \bibinfo{pages}{180502} (\bibinfo{year}{2017}).

\bibitem[{\citenamefont{Kaluarachchi et~al.}(2016)\citenamefont{Kaluarachchi,
  Taufour, B\"ohmer, Tanatar, Bud'ko, Kogan, Prozorov, and
  Canfield}}]{Kaluarachchi16}
\bibinfo{author}{\bibfnamefont{U.~S.} \bibnamefont{Kaluarachchi}},
  \bibinfo{author}{\bibfnamefont{V.}~\bibnamefont{Taufour}},
  \bibinfo{author}{\bibfnamefont{A.~E.} \bibnamefont{B\"ohmer}},
  \bibinfo{author}{\bibfnamefont{M.~A.} \bibnamefont{Tanatar}},
  \bibinfo{author}{\bibfnamefont{S.~L.} \bibnamefont{Bud'ko}},
  \bibinfo{author}{\bibfnamefont{V.~G.} \bibnamefont{Kogan}},
  \bibinfo{author}{\bibfnamefont{R.}~\bibnamefont{Prozorov}}, \bibnamefont{and}
  \bibinfo{author}{\bibfnamefont{P.~C.} \bibnamefont{Canfield}},
  \bibinfo{journal}{Phys. Rev. B} \textbf{\bibinfo{volume}{93}},
  \bibinfo{pages}{064503} (\bibinfo{year}{2016}).

\bibitem[{\citenamefont{Miyoshi et~al.}(2014)\citenamefont{Miyoshi, Morishita,
  Mutou, Kondo, Seida, Fujiwara, Takeuchi, and Nishigori}}]{Miyoshi14}
\bibinfo{author}{\bibfnamefont{K.}~\bibnamefont{Miyoshi}},
  \bibinfo{author}{\bibfnamefont{K.}~\bibnamefont{Morishita}},
  \bibinfo{author}{\bibfnamefont{E.}~\bibnamefont{Mutou}},
  \bibinfo{author}{\bibfnamefont{M.}~\bibnamefont{Kondo}},
  \bibinfo{author}{\bibfnamefont{O.}~\bibnamefont{Seida}},
  \bibinfo{author}{\bibfnamefont{K.}~\bibnamefont{Fujiwara}},
  \bibinfo{author}{\bibfnamefont{J.}~\bibnamefont{Takeuchi}}, \bibnamefont{and}
  \bibinfo{author}{\bibfnamefont{S.}~\bibnamefont{Nishigori}},
  \bibinfo{journal}{Journal of the Physical Society of Japan}
  \textbf{\bibinfo{volume}{83}}, \bibinfo{pages}{013702}
  (\bibinfo{year}{2014}).

\bibitem[{\citenamefont{Kn\"oner et~al.}(2015)\citenamefont{Kn\"oner, Zielke,
  K\"ohler, Wolf, Wolf, Wang, B\"ohmer, Meingast, and Lang}}]{Knoener15}
\bibinfo{author}{\bibfnamefont{S.}~\bibnamefont{Kn\"oner}},
  \bibinfo{author}{\bibfnamefont{D.}~\bibnamefont{Zielke}},
  \bibinfo{author}{\bibfnamefont{S.}~\bibnamefont{K\"ohler}},
  \bibinfo{author}{\bibfnamefont{B.}~\bibnamefont{Wolf}},
  \bibinfo{author}{\bibfnamefont{T.}~\bibnamefont{Wolf}},
  \bibinfo{author}{\bibfnamefont{L.}~\bibnamefont{Wang}},
  \bibinfo{author}{\bibfnamefont{A.}~\bibnamefont{B\"ohmer}},
  \bibinfo{author}{\bibfnamefont{C.}~\bibnamefont{Meingast}}, \bibnamefont{and}
  \bibinfo{author}{\bibfnamefont{M.}~\bibnamefont{Lang}},
  \bibinfo{journal}{Phys. Rev. B} \textbf{\bibinfo{volume}{91}},
  \bibinfo{pages}{174510} (\bibinfo{year}{2015}).

\bibitem[{\citenamefont{Terashima et~al.}(2015)\citenamefont{Terashima,
  Kikugawa, Kasahara, Watashige, Shibauchi, Matsuda, Wolf, Böhmer, Hardy,
  Meingast et~al.}}]{Terashima15}
\bibinfo{author}{\bibfnamefont{T.}~\bibnamefont{Terashima}},
  \bibinfo{author}{\bibfnamefont{N.}~\bibnamefont{Kikugawa}},
  \bibinfo{author}{\bibfnamefont{S.}~\bibnamefont{Kasahara}},
  \bibinfo{author}{\bibfnamefont{T.}~\bibnamefont{Watashige}},
  \bibinfo{author}{\bibfnamefont{T.}~\bibnamefont{Shibauchi}},
  \bibinfo{author}{\bibfnamefont{Y.}~\bibnamefont{Matsuda}},
  \bibinfo{author}{\bibfnamefont{T.}~\bibnamefont{Wolf}},
  \bibinfo{author}{\bibfnamefont{A.~E.} \bibnamefont{Böhmer}},
  \bibinfo{author}{\bibfnamefont{F.}~\bibnamefont{Hardy}},
  \bibinfo{author}{\bibfnamefont{C.}~\bibnamefont{Meingast}},
  \bibnamefont{et~al.}, \bibinfo{journal}{Journal of the Physical Society of
  Japan} \textbf{\bibinfo{volume}{84}}, \bibinfo{pages}{063701}
  (\bibinfo{year}{2015}).

\bibitem[{\citenamefont{Sun et~al.}(2016)\citenamefont{Sun, Matsuura, Ye,
  Mizukami, Shimozawa, Matsubayashi, Yamashita, Watashige, Kasahara, Matsuda
  et~al.}}]{Sun16}
\bibinfo{author}{\bibfnamefont{J.~P.} \bibnamefont{Sun}},
  \bibinfo{author}{\bibfnamefont{K.}~\bibnamefont{Matsuura}},
  \bibinfo{author}{\bibfnamefont{G.~Z.} \bibnamefont{Ye}},
  \bibinfo{author}{\bibfnamefont{Y.}~\bibnamefont{Mizukami}},
  \bibinfo{author}{\bibfnamefont{M.}~\bibnamefont{Shimozawa}},
  \bibinfo{author}{\bibfnamefont{K.}~\bibnamefont{Matsubayashi}},
  \bibinfo{author}{\bibfnamefont{M.}~\bibnamefont{Yamashita}},
  \bibinfo{author}{\bibfnamefont{T.}~\bibnamefont{Watashige}},
  \bibinfo{author}{\bibfnamefont{S.}~\bibnamefont{Kasahara}},
  \bibinfo{author}{\bibfnamefont{Y.}~\bibnamefont{Matsuda}},
  \bibnamefont{et~al.}, \bibinfo{journal}{Nat. Commun.}
  \textbf{\bibinfo{volume}{7}}, \bibinfo{pages}{12146} (\bibinfo{year}{2016}).

\bibitem[{\citenamefont{Bendele et~al.}(2012)\citenamefont{Bendele, Ichsanow,
  Pashkevich, Keller, Str\"assle, Gusev, Pomjakushina, Conder, Khasanov, and
  Keller}}]{Bendele12}
\bibinfo{author}{\bibfnamefont{M.}~\bibnamefont{Bendele}},
  \bibinfo{author}{\bibfnamefont{A.}~\bibnamefont{Ichsanow}},
  \bibinfo{author}{\bibfnamefont{Y.}~\bibnamefont{Pashkevich}},
  \bibinfo{author}{\bibfnamefont{L.}~\bibnamefont{Keller}},
  \bibinfo{author}{\bibfnamefont{T.}~\bibnamefont{Str\"assle}},
  \bibinfo{author}{\bibfnamefont{A.}~\bibnamefont{Gusev}},
  \bibinfo{author}{\bibfnamefont{E.}~\bibnamefont{Pomjakushina}},
  \bibinfo{author}{\bibfnamefont{K.}~\bibnamefont{Conder}},
  \bibinfo{author}{\bibfnamefont{R.}~\bibnamefont{Khasanov}}, \bibnamefont{and}
  \bibinfo{author}{\bibfnamefont{H.}~\bibnamefont{Keller}},
  \bibinfo{journal}{Phys. Rev. B} \textbf{\bibinfo{volume}{85}},
  \bibinfo{pages}{064517} (\bibinfo{year}{2012}).

\bibitem[{\citenamefont{Khasanov et~al.}(2018)\citenamefont{Khasanov,
  Fernandes, Simutis, Guguchia, Amato, Luetkens, Morenzoni, Dong, Zhou, and
  Zhao}}]{Khasanov18}
\bibinfo{author}{\bibfnamefont{R.}~\bibnamefont{Khasanov}},
  \bibinfo{author}{\bibfnamefont{R.~M.} \bibnamefont{Fernandes}},
  \bibinfo{author}{\bibfnamefont{G.}~\bibnamefont{Simutis}},
  \bibinfo{author}{\bibfnamefont{Z.}~\bibnamefont{Guguchia}},
  \bibinfo{author}{\bibfnamefont{A.}~\bibnamefont{Amato}},
  \bibinfo{author}{\bibfnamefont{H.}~\bibnamefont{Luetkens}},
  \bibinfo{author}{\bibfnamefont{E.}~\bibnamefont{Morenzoni}},
  \bibinfo{author}{\bibfnamefont{X.}~\bibnamefont{Dong}},
  \bibinfo{author}{\bibfnamefont{F.}~\bibnamefont{Zhou}}, \bibnamefont{and}
  \bibinfo{author}{\bibfnamefont{Z.}~\bibnamefont{Zhao}},
  \bibinfo{journal}{Phys. Rev. B} \textbf{\bibinfo{volume}{97}},
  \bibinfo{pages}{224510} (\bibinfo{year}{2018}).

\bibitem[{\citenamefont{Böhmer et~al.}(2018)\citenamefont{Böhmer, Kothapalli,
  Jayasekara, Wilde, Li, Sapkota, Ueland, Das, Xiao, Bi et~al.}}]{Boehmer18}
\bibinfo{author}{\bibfnamefont{A.~E.} \bibnamefont{Böhmer}},
  \bibinfo{author}{\bibfnamefont{K.}~\bibnamefont{Kothapalli}},
  \bibinfo{author}{\bibfnamefont{W.~T.} \bibnamefont{Jayasekara}},
  \bibinfo{author}{\bibfnamefont{J.~M.} \bibnamefont{Wilde}},
  \bibinfo{author}{\bibfnamefont{B.}~\bibnamefont{Li}},
  \bibinfo{author}{\bibfnamefont{A.}~\bibnamefont{Sapkota}},
  \bibinfo{author}{\bibfnamefont{B.~G.} \bibnamefont{Ueland}},
  \bibinfo{author}{\bibfnamefont{P.}~\bibnamefont{Das}},
  \bibinfo{author}{\bibfnamefont{Y.}~\bibnamefont{Xiao}},
  \bibinfo{author}{\bibfnamefont{W.}~\bibnamefont{Bi}}, \bibnamefont{et~al.},
  \bibinfo{journal}{arXiv} p. \bibinfo{pages}{1803.09449}
  (\bibinfo{year}{2018}).

\bibitem[{\citenamefont{Svitlyk et~al.}(2017)\citenamefont{Svitlyk, Raba,
  Dmitriev, Rodi\`ere, Toulemonde, Chernyshov, and Mezouar}}]{Svitlyk17}
\bibinfo{author}{\bibfnamefont{V.}~\bibnamefont{Svitlyk}},
  \bibinfo{author}{\bibfnamefont{M.}~\bibnamefont{Raba}},
  \bibinfo{author}{\bibfnamefont{V.}~\bibnamefont{Dmitriev}},
  \bibinfo{author}{\bibfnamefont{P.}~\bibnamefont{Rodi\`ere}},
  \bibinfo{author}{\bibfnamefont{P.}~\bibnamefont{Toulemonde}},
  \bibinfo{author}{\bibfnamefont{D.}~\bibnamefont{Chernyshov}},
  \bibnamefont{and} \bibinfo{author}{\bibfnamefont{M.}~\bibnamefont{Mezouar}},
  \bibinfo{journal}{Phys. Rev. B} \textbf{\bibinfo{volume}{96}},
  \bibinfo{pages}{014520} (\bibinfo{year}{2017}).

\bibitem[{\citenamefont{Lebert et~al.}(2018)\citenamefont{Lebert, Bal\'edent,
  Toulemonde, Ablett, and Rueff}}]{Lebert18}
\bibinfo{author}{\bibfnamefont{B.~W.} \bibnamefont{Lebert}},
  \bibinfo{author}{\bibfnamefont{V.}~\bibnamefont{Bal\'edent}},
  \bibinfo{author}{\bibfnamefont{P.}~\bibnamefont{Toulemonde}},
  \bibinfo{author}{\bibfnamefont{J.~M.} \bibnamefont{Ablett}},
  \bibnamefont{and} \bibinfo{author}{\bibfnamefont{J.-P.} \bibnamefont{Rueff}},
  \bibinfo{journal}{Phys. Rev. B} \textbf{\bibinfo{volume}{97}},
  \bibinfo{pages}{180503} (\bibinfo{year}{2018}).

\bibitem[{\citenamefont{Khasanov et~al.}(2017)\citenamefont{Khasanov, Guguchia,
  Amato, Morenzoni, Dong, Zhou, and Zhao}}]{Khasanov17}
\bibinfo{author}{\bibfnamefont{R.}~\bibnamefont{Khasanov}},
  \bibinfo{author}{\bibfnamefont{Z.}~\bibnamefont{Guguchia}},
  \bibinfo{author}{\bibfnamefont{A.}~\bibnamefont{Amato}},
  \bibinfo{author}{\bibfnamefont{E.}~\bibnamefont{Morenzoni}},
  \bibinfo{author}{\bibfnamefont{X.}~\bibnamefont{Dong}},
  \bibinfo{author}{\bibfnamefont{F.}~\bibnamefont{Zhou}}, \bibnamefont{and}
  \bibinfo{author}{\bibfnamefont{Z.}~\bibnamefont{Zhao}},
  \bibinfo{journal}{Phys. Rev. B} \textbf{\bibinfo{volume}{95}},
  \bibinfo{pages}{180504} (\bibinfo{year}{2017}).

\bibitem[{\citenamefont{Chubukov et~al.}(2016)\citenamefont{Chubukov, Khodas,
  and Fernandes}}]{Fernandes16}
\bibinfo{author}{\bibfnamefont{A.~V.} \bibnamefont{Chubukov}},
  \bibinfo{author}{\bibfnamefont{M.}~\bibnamefont{Khodas}}, \bibnamefont{and}
  \bibinfo{author}{\bibfnamefont{R.~M.} \bibnamefont{Fernandes}},
  \bibinfo{journal}{Phys. Rev. X} \textbf{\bibinfo{volume}{6}},
  \bibinfo{pages}{041045} (\bibinfo{year}{2016}).

\bibitem[{\citenamefont{Glasbrenner et~al.}(2015)\citenamefont{Glasbrenner,
  Mazin, Jeschke, P.~J.~Hirschfeld, and Valentí}}]{Glasbrenner15}
\bibinfo{author}{\bibfnamefont{J.~K.} \bibnamefont{Glasbrenner}},
  \bibinfo{author}{\bibfnamefont{I.~I.} \bibnamefont{Mazin}},
  \bibinfo{author}{\bibfnamefont{H.~O.} \bibnamefont{Jeschke}},
  \bibinfo{author}{\bibfnamefont{R.~M.~F.} \bibnamefont{P.~J.~Hirschfeld}},
  \bibnamefont{and} \bibinfo{author}{\bibfnamefont{R.}~\bibnamefont{Valentí}},
  \bibinfo{journal}{Nature Physics} \textbf{\bibinfo{volume}{11}},
  \bibinfo{pages}{953–958} (\bibinfo{year}{2015}).

\bibitem[{\citenamefont{Onari et~al.}(2016)\citenamefont{Onari, Yamakawa, and
  Kontani}}]{Onari16}
\bibinfo{author}{\bibfnamefont{S.}~\bibnamefont{Onari}},
  \bibinfo{author}{\bibfnamefont{Y.}~\bibnamefont{Yamakawa}}, \bibnamefont{and}
  \bibinfo{author}{\bibfnamefont{H.}~\bibnamefont{Kontani}},
  \bibinfo{journal}{Phys. Rev. Lett.} \textbf{\bibinfo{volume}{116}},
  \bibinfo{pages}{227001} (\bibinfo{year}{2016}).

\bibitem[{\citenamefont{Yamakawa et~al.}(2016)\citenamefont{Yamakawa, Onari,
  and Kontani}}]{Yamakawa16}
\bibinfo{author}{\bibfnamefont{Y.}~\bibnamefont{Yamakawa}},
  \bibinfo{author}{\bibfnamefont{S.}~\bibnamefont{Onari}}, \bibnamefont{and}
  \bibinfo{author}{\bibfnamefont{H.}~\bibnamefont{Kontani}},
  \bibinfo{journal}{Phys. Rev. X} \textbf{\bibinfo{volume}{6}},
  \bibinfo{pages}{021032} (\bibinfo{year}{2016}).

\bibitem[{\citenamefont{Chubukov et~al.}(2015)\citenamefont{Chubukov,
  Fernandes, and Schmalian}}]{Chubukov15}
\bibinfo{author}{\bibfnamefont{A.~V.} \bibnamefont{Chubukov}},
  \bibinfo{author}{\bibfnamefont{R.~M.} \bibnamefont{Fernandes}},
  \bibnamefont{and}
  \bibinfo{author}{\bibfnamefont{J.}~\bibnamefont{Schmalian}},
  \bibinfo{journal}{Phys. Rev. B} \textbf{\bibinfo{volume}{91}},
  \bibinfo{pages}{201105} (\bibinfo{year}{2015}).

\bibitem[{\citenamefont{Yu and Si}(2015)}]{Yu15}
\bibinfo{author}{\bibfnamefont{R.}~\bibnamefont{Yu}} \bibnamefont{and}
  \bibinfo{author}{\bibfnamefont{Q.}~\bibnamefont{Si}}, \bibinfo{journal}{Phys.
  Rev. Lett.} \textbf{\bibinfo{volume}{115}}, \bibinfo{pages}{116401}
  (\bibinfo{year}{2015}).

\bibitem[{\citenamefont{Wang et~al.}(2015)\citenamefont{Wang, Kivelson, and
  Lee}}]{Wang15}
\bibinfo{author}{\bibfnamefont{F.}~\bibnamefont{Wang}},
  \bibinfo{author}{\bibfnamefont{S.~A.} \bibnamefont{Kivelson}},
  \bibnamefont{and} \bibinfo{author}{\bibfnamefont{D.-H.} \bibnamefont{Lee}},
  \bibinfo{journal}{Nature Physics} \textbf{\bibinfo{volume}{11}},
  \bibinfo{pages}{959–963} (\bibinfo{year}{2015}).

\bibitem[{\citenamefont{Chen et~al.}(2019)\citenamefont{Chen, Wang, Zhu, and
  Wen}}]{Chen19}
\bibinfo{author}{\bibfnamefont{G.-Y.} \bibnamefont{Chen}},
  \bibinfo{author}{\bibfnamefont{E.}~\bibnamefont{Wang}},
  \bibinfo{author}{\bibfnamefont{X.}~\bibnamefont{Zhu}}, \bibnamefont{and}
  \bibinfo{author}{\bibfnamefont{H.-H.} \bibnamefont{Wen}},
  \bibinfo{journal}{Phys. Rev. B} \textbf{\bibinfo{volume}{99}},
  \bibinfo{pages}{054517} (\bibinfo{year}{2019}).

\bibitem[{\citenamefont{Yip et~al.}(2017)\citenamefont{Yip, Chan, Niu,
  Matsuura, Mizukami, Kasahara, Matsuda, Shibauchi, and Goh}}]{Yip17}
\bibinfo{author}{\bibfnamefont{K.~Y.} \bibnamefont{Yip}},
  \bibinfo{author}{\bibfnamefont{Y.~C.} \bibnamefont{Chan}},
  \bibinfo{author}{\bibfnamefont{Q.}~\bibnamefont{Niu}},
  \bibinfo{author}{\bibfnamefont{K.}~\bibnamefont{Matsuura}},
  \bibinfo{author}{\bibfnamefont{Y.}~\bibnamefont{Mizukami}},
  \bibinfo{author}{\bibfnamefont{S.}~\bibnamefont{Kasahara}},
  \bibinfo{author}{\bibfnamefont{Y.}~\bibnamefont{Matsuda}},
  \bibinfo{author}{\bibfnamefont{T.}~\bibnamefont{Shibauchi}},
  \bibnamefont{and} \bibinfo{author}{\bibfnamefont{S.~K.} \bibnamefont{Goh}},
  \bibinfo{journal}{Phys. Rev. B} \textbf{\bibinfo{volume}{96}},
  \bibinfo{pages}{020502} (\bibinfo{year}{2017}).

\bibitem[{\citenamefont{Gati et~al.}(2019)\citenamefont{Gati, Drachuck, Xiang,
  Wang, Bud'ko, and Canfield}}]{Gati19}
\bibinfo{author}{\bibfnamefont{E.}~\bibnamefont{Gati}},
  \bibinfo{author}{\bibfnamefont{G.}~\bibnamefont{Drachuck}},
  \bibinfo{author}{\bibfnamefont{L.}~\bibnamefont{Xiang}},
  \bibinfo{author}{\bibfnamefont{L.-L.} \bibnamefont{Wang}},
  \bibinfo{author}{\bibfnamefont{S.~L.} \bibnamefont{Bud'ko}},
  \bibnamefont{and} \bibinfo{author}{\bibfnamefont{P.~C.}
  \bibnamefont{Canfield}}, \bibinfo{journal}{Review of Scientific Instruments}
  \textbf{\bibinfo{volume}{90}}, \bibinfo{pages}{023911}
  (\bibinfo{year}{2019}).

\bibitem[{\citenamefont{B\"ohmer et~al.}(2016)\citenamefont{B\"ohmer, Taufour,
  Straszheim, Wolf, and Canfield}}]{Boehmer16}
\bibinfo{author}{\bibfnamefont{A.~E.} \bibnamefont{B\"ohmer}},
  \bibinfo{author}{\bibfnamefont{V.}~\bibnamefont{Taufour}},
  \bibinfo{author}{\bibfnamefont{W.~E.} \bibnamefont{Straszheim}},
  \bibinfo{author}{\bibfnamefont{T.}~\bibnamefont{Wolf}}, \bibnamefont{and}
  \bibinfo{author}{\bibfnamefont{P.~C.} \bibnamefont{Canfield}},
  \bibinfo{journal}{Phys. Rev. B} \textbf{\bibinfo{volume}{94}},
  \bibinfo{pages}{024526} (\bibinfo{year}{2016}).

\bibitem[{\citenamefont{Torikachvili et~al.}(2015)\citenamefont{Torikachvili,
  Kim, Colombier, Bud’ko, and Canfield}}]{Torikachvili15}
\bibinfo{author}{\bibfnamefont{M.~S.} \bibnamefont{Torikachvili}},
  \bibinfo{author}{\bibfnamefont{S.~K.} \bibnamefont{Kim}},
  \bibinfo{author}{\bibfnamefont{E.}~\bibnamefont{Colombier}},
  \bibinfo{author}{\bibfnamefont{S.~L.} \bibnamefont{Bud’ko}},
  \bibnamefont{and} \bibinfo{author}{\bibfnamefont{P.~C.}
  \bibnamefont{Canfield}}, \bibinfo{journal}{Review of Scientific Instruments}
  \textbf{\bibinfo{volume}{86}}, \bibinfo{pages}{123904}
  (\bibinfo{year}{2015}).

\bibitem[{\citenamefont{Cheung et~al.}(2018)\citenamefont{Cheung, Guguchia,
  Frandsen, Gong, Yamakawa, Almeida, Onuorah, Bonf\'a, Miranda, Wang
  et~al.}}]{Cheung18}
\bibinfo{author}{\bibfnamefont{S.~C.} \bibnamefont{Cheung}},
  \bibinfo{author}{\bibfnamefont{Z.}~\bibnamefont{Guguchia}},
  \bibinfo{author}{\bibfnamefont{B.~A.} \bibnamefont{Frandsen}},
  \bibinfo{author}{\bibfnamefont{Z.}~\bibnamefont{Gong}},
  \bibinfo{author}{\bibfnamefont{K.}~\bibnamefont{Yamakawa}},
  \bibinfo{author}{\bibfnamefont{D.~E.} \bibnamefont{Almeida}},
  \bibinfo{author}{\bibfnamefont{I.~J.} \bibnamefont{Onuorah}},
  \bibinfo{author}{\bibfnamefont{P.}~\bibnamefont{Bonf\'a}},
  \bibinfo{author}{\bibfnamefont{E.}~\bibnamefont{Miranda}},
  \bibinfo{author}{\bibfnamefont{W.}~\bibnamefont{Wang}}, \bibnamefont{et~al.},
  \bibinfo{journal}{Phys. Rev. B} \textbf{\bibinfo{volume}{97}},
  \bibinfo{pages}{224508} (\bibinfo{year}{2018}).

\bibitem[{\citenamefont{Machida}(1981)}]{Machida81}
\bibinfo{author}{\bibfnamefont{K.}~\bibnamefont{Machida}},
  \bibinfo{journal}{Journal of the Physical Society of Japan}
  \textbf{\bibinfo{volume}{50}}, \bibinfo{pages}{2195} (\bibinfo{year}{1981}).

\bibitem[{\citenamefont{Rotter et~al.}(2009)\citenamefont{Rotter, Tegel,
  Schellenberg, Schappacher, Pöttgen, Deisenhofer, Günther, Schrettle, Loidl,
  and Johrendt}}]{Rotter09}
\bibinfo{author}{\bibfnamefont{M.}~\bibnamefont{Rotter}},
  \bibinfo{author}{\bibfnamefont{M.}~\bibnamefont{Tegel}},
  \bibinfo{author}{\bibfnamefont{I.}~\bibnamefont{Schellenberg}},
  \bibinfo{author}{\bibfnamefont{F.~M.} \bibnamefont{Schappacher}},
  \bibinfo{author}{\bibfnamefont{R.}~\bibnamefont{Pöttgen}},
  \bibinfo{author}{\bibfnamefont{J.}~\bibnamefont{Deisenhofer}},
  \bibinfo{author}{\bibfnamefont{A.}~\bibnamefont{Günther}},
  \bibinfo{author}{\bibfnamefont{F.}~\bibnamefont{Schrettle}},
  \bibinfo{author}{\bibfnamefont{A.}~\bibnamefont{Loidl}}, \bibnamefont{and}
  \bibinfo{author}{\bibfnamefont{D.}~\bibnamefont{Johrendt}},
  \bibinfo{journal}{New Journal of Physics} \textbf{\bibinfo{volume}{11}},
  \bibinfo{pages}{025014} (\bibinfo{year}{2009}).

\bibitem[{\citenamefont{Pratt et~al.}(2009)\citenamefont{Pratt, Tian, Kreyssig,
  Zarestky, Nandi, Ni, Bud'ko, Canfield, Goldman, and McQueeney}}]{Pratt09}
\bibinfo{author}{\bibfnamefont{D.~K.} \bibnamefont{Pratt}},
  \bibinfo{author}{\bibfnamefont{W.}~\bibnamefont{Tian}},
  \bibinfo{author}{\bibfnamefont{A.}~\bibnamefont{Kreyssig}},
  \bibinfo{author}{\bibfnamefont{J.~L.} \bibnamefont{Zarestky}},
  \bibinfo{author}{\bibfnamefont{S.}~\bibnamefont{Nandi}},
  \bibinfo{author}{\bibfnamefont{N.}~\bibnamefont{Ni}},
  \bibinfo{author}{\bibfnamefont{S.~L.} \bibnamefont{Bud'ko}},
  \bibinfo{author}{\bibfnamefont{P.~C.} \bibnamefont{Canfield}},
  \bibinfo{author}{\bibfnamefont{A.~I.} \bibnamefont{Goldman}},
  \bibnamefont{and} \bibinfo{author}{\bibfnamefont{R.~J.}
  \bibnamefont{McQueeney}}, \bibinfo{journal}{Phys. Rev. Lett.}
  \textbf{\bibinfo{volume}{103}}, \bibinfo{pages}{087001}
  (\bibinfo{year}{2009}).

\bibitem[{\citenamefont{Christianson et~al.}(2009)\citenamefont{Christianson,
  Lumsden, Nagler, MacDougall, McGuire, Sefat, Jin, Sales, and
  Mandrus}}]{Christianson09}
\bibinfo{author}{\bibfnamefont{A.~D.} \bibnamefont{Christianson}},
  \bibinfo{author}{\bibfnamefont{M.~D.} \bibnamefont{Lumsden}},
  \bibinfo{author}{\bibfnamefont{S.~E.} \bibnamefont{Nagler}},
  \bibinfo{author}{\bibfnamefont{G.~J.} \bibnamefont{MacDougall}},
  \bibinfo{author}{\bibfnamefont{M.~A.} \bibnamefont{McGuire}},
  \bibinfo{author}{\bibfnamefont{A.~S.} \bibnamefont{Sefat}},
  \bibinfo{author}{\bibfnamefont{R.}~\bibnamefont{Jin}},
  \bibinfo{author}{\bibfnamefont{B.~C.} \bibnamefont{Sales}}, \bibnamefont{and}
  \bibinfo{author}{\bibfnamefont{D.}~\bibnamefont{Mandrus}},
  \bibinfo{journal}{Phys. Rev. Lett.} \textbf{\bibinfo{volume}{103}},
  \bibinfo{pages}{087002} (\bibinfo{year}{2009}).

\bibitem[{\citenamefont{Luo et~al.}(2012)\citenamefont{Luo, Zhang, Laver,
  Yamani, Wang, Lu, Wang, Chen, Li, Chang et~al.}}]{Luo12}
\bibinfo{author}{\bibfnamefont{H.}~\bibnamefont{Luo}},
  \bibinfo{author}{\bibfnamefont{R.}~\bibnamefont{Zhang}},
  \bibinfo{author}{\bibfnamefont{M.}~\bibnamefont{Laver}},
  \bibinfo{author}{\bibfnamefont{Z.}~\bibnamefont{Yamani}},
  \bibinfo{author}{\bibfnamefont{M.}~\bibnamefont{Wang}},
  \bibinfo{author}{\bibfnamefont{X.}~\bibnamefont{Lu}},
  \bibinfo{author}{\bibfnamefont{M.}~\bibnamefont{Wang}},
  \bibinfo{author}{\bibfnamefont{Y.}~\bibnamefont{Chen}},
  \bibinfo{author}{\bibfnamefont{S.}~\bibnamefont{Li}},
  \bibinfo{author}{\bibfnamefont{S.}~\bibnamefont{Chang}},
  \bibnamefont{et~al.}, \bibinfo{journal}{Phys. Rev. Lett.}
  \textbf{\bibinfo{volume}{108}}, \bibinfo{pages}{247002}
  (\bibinfo{year}{2012}).

\bibitem[{\citenamefont{Fernandes and Schmalian}(2010)}]{Fernandes10}
\bibinfo{author}{\bibfnamefont{R.~M.} \bibnamefont{Fernandes}}
  \bibnamefont{and}
  \bibinfo{author}{\bibfnamefont{J.}~\bibnamefont{Schmalian}},
  \bibinfo{journal}{Phys. Rev. B} \textbf{\bibinfo{volume}{82}},
  \bibinfo{pages}{014521} (\bibinfo{year}{2010}).

\bibitem[{\citenamefont{Böhmer et~al.}(2015)\citenamefont{Böhmer, Hardy,
  Wang, Wolf, Schweiss, and Meingast}}]{Boehmer15b}
\bibinfo{author}{\bibfnamefont{A.~E.} \bibnamefont{Böhmer}},
  \bibinfo{author}{\bibfnamefont{F.}~\bibnamefont{Hardy}},
  \bibinfo{author}{\bibfnamefont{L.}~\bibnamefont{Wang}},
  \bibinfo{author}{\bibfnamefont{T.}~\bibnamefont{Wolf}},
  \bibinfo{author}{\bibfnamefont{P.}~\bibnamefont{Schweiss}}, \bibnamefont{and}
  \bibinfo{author}{\bibfnamefont{C.}~\bibnamefont{Meingast}},
  \bibinfo{journal}{Nature Communications} \textbf{\bibinfo{volume}{6}},
  \bibinfo{pages}{7911} (\bibinfo{year}{2015}).

\bibitem[{\citenamefont{Bud'ko et~al.}(2018)\citenamefont{Bud'ko, Kogan,
  Prozorov, Meier, Xu, and Canfield}}]{Budko18}
\bibinfo{author}{\bibfnamefont{S.~L.} \bibnamefont{Bud'ko}},
  \bibinfo{author}{\bibfnamefont{V.~G.} \bibnamefont{Kogan}},
  \bibinfo{author}{\bibfnamefont{R.}~\bibnamefont{Prozorov}},
  \bibinfo{author}{\bibfnamefont{W.~R.} \bibnamefont{Meier}},
  \bibinfo{author}{\bibfnamefont{M.}~\bibnamefont{Xu}}, \bibnamefont{and}
  \bibinfo{author}{\bibfnamefont{P.~C.} \bibnamefont{Canfield}},
  \bibinfo{journal}{Phys. Rev. B} \textbf{\bibinfo{volume}{98}},
  \bibinfo{pages}{144520} (\bibinfo{year}{2018}).

\bibitem[{foo()}]{footnote}
\bibinfo{note}{We omit data points of Refs. \cite{Wang16,Sun16} due to pressure
  offsets, and Ref. \cite{Miyoshi14} due to a different criterion for $T_s$.}

\bibitem[{\citenamefont{Junod et~al.}(1999)\citenamefont{Junod, Erb, and
  Renner}}]{Junod99}
\bibinfo{author}{\bibfnamefont{A.}~\bibnamefont{Junod}},
  \bibinfo{author}{\bibfnamefont{A.}~\bibnamefont{Erb}}, \bibnamefont{and}
  \bibinfo{author}{\bibfnamefont{C.}~\bibnamefont{Renner}},
  \bibinfo{journal}{Physica C: Superconductivity}
  \textbf{\bibinfo{volume}{317-318}}, \bibinfo{pages}{333 }
  (\bibinfo{year}{1999}).

\bibitem[{\citenamefont{Adachi and Ikeda}(2017)}]{Adachi17}
\bibinfo{author}{\bibfnamefont{K.}~\bibnamefont{Adachi}} \bibnamefont{and}
  \bibinfo{author}{\bibfnamefont{R.}~\bibnamefont{Ikeda}},
  \bibinfo{journal}{Phys. Rev. B} \textbf{\bibinfo{volume}{96}},
  \bibinfo{pages}{184507} (\bibinfo{year}{2017}).

\bibitem[{\citenamefont{Yang et~al.}(2017)\citenamefont{Yang, Chen, Zhu, Xing,
  and Wen}}]{Yang17}
\bibinfo{author}{\bibfnamefont{H.}~\bibnamefont{Yang}},
  \bibinfo{author}{\bibfnamefont{G.}~\bibnamefont{Chen}},
  \bibinfo{author}{\bibfnamefont{X.}~\bibnamefont{Zhu}},
  \bibinfo{author}{\bibfnamefont{J.}~\bibnamefont{Xing}}, \bibnamefont{and}
  \bibinfo{author}{\bibfnamefont{H.-H.} \bibnamefont{Wen}},
  \bibinfo{journal}{Phys. Rev. B} \textbf{\bibinfo{volume}{96}},
  \bibinfo{pages}{064501} (\bibinfo{year}{2017}).

\bibitem[{\citenamefont{Terashima et~al.}(2016)\citenamefont{Terashima,
  Kikugawa, Kiswandhi, Graf, Choi, Brooks, Kasahara, Watashige, Matsuda,
  Shibauchi et~al.}}]{Terashima16}
\bibinfo{author}{\bibfnamefont{T.}~\bibnamefont{Terashima}},
  \bibinfo{author}{\bibfnamefont{N.}~\bibnamefont{Kikugawa}},
  \bibinfo{author}{\bibfnamefont{A.}~\bibnamefont{Kiswandhi}},
  \bibinfo{author}{\bibfnamefont{D.}~\bibnamefont{Graf}},
  \bibinfo{author}{\bibfnamefont{E.-S.} \bibnamefont{Choi}},
  \bibinfo{author}{\bibfnamefont{J.~S.} \bibnamefont{Brooks}},
  \bibinfo{author}{\bibfnamefont{S.}~\bibnamefont{Kasahara}},
  \bibinfo{author}{\bibfnamefont{T.}~\bibnamefont{Watashige}},
  \bibinfo{author}{\bibfnamefont{Y.}~\bibnamefont{Matsuda}},
  \bibinfo{author}{\bibfnamefont{T.}~\bibnamefont{Shibauchi}},
  \bibnamefont{et~al.}, \bibinfo{journal}{Phys. Rev. B}
  \textbf{\bibinfo{volume}{93}}, \bibinfo{pages}{094505}
  (\bibinfo{year}{2016}).

\bibitem[{\citenamefont{Massat et~al.}(2018)\citenamefont{Massat, Quan,
  Grasset, M\'easson, Cazayous, Sacuto, Karlsson, Strobel, Toulemonde, Yin
  et~al.}}]{Massat18}
\bibinfo{author}{\bibfnamefont{P.}~\bibnamefont{Massat}},
  \bibinfo{author}{\bibfnamefont{Y.}~\bibnamefont{Quan}},
  \bibinfo{author}{\bibfnamefont{R.}~\bibnamefont{Grasset}},
  \bibinfo{author}{\bibfnamefont{M.-A.} \bibnamefont{M\'easson}},
  \bibinfo{author}{\bibfnamefont{M.}~\bibnamefont{Cazayous}},
  \bibinfo{author}{\bibfnamefont{A.}~\bibnamefont{Sacuto}},
  \bibinfo{author}{\bibfnamefont{S.}~\bibnamefont{Karlsson}},
  \bibinfo{author}{\bibfnamefont{P.}~\bibnamefont{Strobel}},
  \bibinfo{author}{\bibfnamefont{P.}~\bibnamefont{Toulemonde}},
  \bibinfo{author}{\bibfnamefont{Z.}~\bibnamefont{Yin}}, \bibnamefont{et~al.},
  \bibinfo{journal}{Phys. Rev. Lett.} \textbf{\bibinfo{volume}{121}},
  \bibinfo{pages}{077001} (\bibinfo{year}{2018}).

\bibitem[{\citenamefont{Martiny et~al.}(2019)\citenamefont{Martiny, Kreisel,
  and Andersen}}]{Martiny19}
\bibinfo{author}{\bibfnamefont{J.~H.~J.} \bibnamefont{Martiny}},
  \bibinfo{author}{\bibfnamefont{A.}~\bibnamefont{Kreisel}}, \bibnamefont{and}
  \bibinfo{author}{\bibfnamefont{B.~M.} \bibnamefont{Andersen}},
  \bibinfo{journal}{Phys. Rev. B} \textbf{\bibinfo{volume}{99}},
  \bibinfo{pages}{014509} (\bibinfo{year}{2019}).

\bibitem[{\citenamefont{Yu and Kivelson}(2019)}]{Yu19}
\bibinfo{author}{\bibfnamefont{Y.}~\bibnamefont{Yu}} \bibnamefont{and}
  \bibinfo{author}{\bibfnamefont{S.~A.} \bibnamefont{Kivelson}},
  \bibinfo{journal}{Phys. Rev. B} \textbf{\bibinfo{volume}{99}},
  \bibinfo{pages}{144513} (\bibinfo{year}{2019}).

\bibitem[{\citenamefont{Keimer et~al.}(2015)\citenamefont{Keimer, Kivelson,
  Norman, Uchida, and Zaanen}}]{Keimer15}
\bibinfo{author}{\bibfnamefont{B.}~\bibnamefont{Keimer}},
  \bibinfo{author}{\bibfnamefont{S.~A.} \bibnamefont{Kivelson}},
  \bibinfo{author}{\bibfnamefont{M.~R.} \bibnamefont{Norman}},
  \bibinfo{author}{\bibfnamefont{S.}~\bibnamefont{Uchida}}, \bibnamefont{and}
  \bibinfo{author}{\bibfnamefont{J.}~\bibnamefont{Zaanen}},
  \bibinfo{journal}{Nature} \textbf{\bibinfo{volume}{518}},
  \bibinfo{pages}{179–186} (\bibinfo{year}{2015}).

\bibitem[{\citenamefont{da~Silva~Neto et~al.}(2014)\citenamefont{da~Silva~Neto,
  Aynajian, Frano, Comin, Schierle, Weschke, Gyenis, Wen, Schneeloch, Xu
  et~al.}}]{SilvaNeto14}
\bibinfo{author}{\bibfnamefont{E.~H.} \bibnamefont{da~Silva~Neto}},
  \bibinfo{author}{\bibfnamefont{P.}~\bibnamefont{Aynajian}},
  \bibinfo{author}{\bibfnamefont{A.}~\bibnamefont{Frano}},
  \bibinfo{author}{\bibfnamefont{R.}~\bibnamefont{Comin}},
  \bibinfo{author}{\bibfnamefont{E.}~\bibnamefont{Schierle}},
  \bibinfo{author}{\bibfnamefont{E.}~\bibnamefont{Weschke}},
  \bibinfo{author}{\bibfnamefont{A.}~\bibnamefont{Gyenis}},
  \bibinfo{author}{\bibfnamefont{J.}~\bibnamefont{Wen}},
  \bibinfo{author}{\bibfnamefont{J.}~\bibnamefont{Schneeloch}},
  \bibinfo{author}{\bibfnamefont{Z.}~\bibnamefont{Xu}}, \bibnamefont{et~al.},
  \bibinfo{journal}{Science} \textbf{\bibinfo{volume}{343}},
  \bibinfo{pages}{393} (\bibinfo{year}{2014}).

\bibitem[{\citenamefont{Kivelson et~al.}(2003)\citenamefont{Kivelson, Bindloss,
  Fradkin, Oganesyan, Tranquada, Kapitulnik, and Howald}}]{Kivelson03}
\bibinfo{author}{\bibfnamefont{S.~A.} \bibnamefont{Kivelson}},
  \bibinfo{author}{\bibfnamefont{I.~P.} \bibnamefont{Bindloss}},
  \bibinfo{author}{\bibfnamefont{E.}~\bibnamefont{Fradkin}},
  \bibinfo{author}{\bibfnamefont{V.}~\bibnamefont{Oganesyan}},
  \bibinfo{author}{\bibfnamefont{J.~M.} \bibnamefont{Tranquada}},
  \bibinfo{author}{\bibfnamefont{A.}~\bibnamefont{Kapitulnik}},
  \bibnamefont{and} \bibinfo{author}{\bibfnamefont{C.}~\bibnamefont{Howald}},
  \bibinfo{journal}{Rev. Mod. Phys.} \textbf{\bibinfo{volume}{75}},
  \bibinfo{pages}{1201} (\bibinfo{year}{2003}).

\bibitem[{\citenamefont{Torchinsky et~al.}(2013)\citenamefont{Torchinsky,
  Mahmood, Bollinger, Božović, and Gedik}}]{Torchinsky13}
\bibinfo{author}{\bibfnamefont{D.~H.} \bibnamefont{Torchinsky}},
  \bibinfo{author}{\bibfnamefont{F.}~\bibnamefont{Mahmood}},
  \bibinfo{author}{\bibfnamefont{A.~T.} \bibnamefont{Bollinger}},
  \bibinfo{author}{\bibfnamefont{I.}~\bibnamefont{Božović}},
  \bibnamefont{and} \bibinfo{author}{\bibfnamefont{N.}~\bibnamefont{Gedik}},
  \bibinfo{journal}{Nature Materials} \textbf{\bibinfo{volume}{12}},
  \bibinfo{pages}{387–391} (\bibinfo{year}{2013}).

\bibitem[{\citenamefont{Vishik}(2018)}]{Vishik18}
\bibinfo{author}{\bibfnamefont{I.~M.} \bibnamefont{Vishik}},
  \bibinfo{journal}{Reports on Progress in Physics}
  \textbf{\bibinfo{volume}{81}}, \bibinfo{pages}{062501}
  (\bibinfo{year}{2018}).

\end{thebibliography}

	\clearpage
	

\section*{Supplementary material}

# Supplementary Information: Bulk superconductivity and role of fluctuations in the iron-based superconductor FeSe at high pressures

Elena Gati[1,2], Anna E. Böhmer[1,2,⋆], Sergey L. Bud'ko[1,2], and Paul C. Canfield[1,2]

[1] *Ames Laboratory, US Department of Energy, Iowa State University, Ames, Iowa 50011, USA*

[2] *Department of Physics and Astronomy, Iowa State University, Ames, Iowa 50011, USA and*

⋆ *current address: Institute for Solid State Physics,*

*Karlsruhe Institute of Technology, 76021 Karlsruhe, Germany*

(Dated: September 10, 2019)

## I. METHODS

Single crystals of FeSe were grown using a modified chemical-vapor transport technique, as described in Ref. 1. The crystal for specific heat measurements under pressure had a plate-like shape (dimensions $\approx 1.5 \times 1 \times 0.5\,\mathrm{mm}^3$) and a mass of about $\sim 1\,\mathrm{mg}$. In Ref. 1, it was reported that the ambient-pressure $T_c$ and $T_s$ of FeSe can vary upon small modifications of the growth procedure. Importantly, the variation of $T_c$ and $T_s$ was found to be correlated with the residual resistivity ratio (RRR) and therefore is considered as an indication for the sample quality. Following these arguments, we chose a crystal with high $T_c$ and $T_s$ for our study. With this being said, it is important to point out that we use hydrostatic pressure as a tuning parameter here on one single sample. Thus, we can exclude that sample-to-sample dependencies (likely due to slightly different stoichometries and varying disorder levels) affect our conclusions.

Our specific heat setup is described in great detail in Ref. 2. In the following, we recall the main aspects which are important for the present work.

Specific heat is measured using the ac calorimetry technique in which the sample is heated in an oscillatory manner and the resulting temperature oscillation contains the information on the specific heat of the sample. This technique has proven to be suited for the use in pressure cells. In general, measurements of the specific heat under pressure do not permit to obtain specific heat values with high absolute accuracy, due to the finite coupling of the sample to the bath. Nevertheless, in our previous work[2], we were able to demonstrate that we can reliably determine changes of the specific heat value by a careful choice of the measurement frequency. As a consequence, the ambient-pressure value of the superconducting jump size, obtained here, $\Delta C_{sc} \approx 79\,\mathrm{mJ/mol/K}$ is different than those of other high-accuracy measurements (see e.g. following values of recent works for comparison: $\Delta C_{sc} \approx 109\,\mathrm{mJ/mol/K}$ (Ref. 3), $97\,\mathrm{mJ/mol/K}$ (Ref. 4), $100\,\mathrm{mJ/mol/K}$ (Ref. 5)). Nevertheless, the strong change of the superconducting jump size $\Delta C_{sc}$ at $T_c$ as a function of pressure can definitely be considered reliable.

A mixture of 4:6 light mineral oil:n-pentane is used as a pressure-transmitting medium. It solidifies at $p \approx 3$-4 GPa at room temperature[6], thus ensuring hydrostatic pressure conditions in the available pressure range[7]. Pressure values, given in the entire manuscript, correspond to those determined from the superconducting critical temperature of elemental lead (Pb)[8], determined resistively.

## II. SPECIFIC HEAT DATA

### A. Detailed view on the phase diagram, determined from specific heat, and comparison with previously-published phase diagrams

Figure S1 shows the temperature-pressure phase diagram of FeSe, determined in the present study from specific heat measurements, in different representations. In Fig. S1 (a), we show the phase diagram which is solely based on our specific heat data. In Fig. S1 (b), we add the structural and magnetic transition temperatures, determined from x-ray diffraction[9] on crystals from the same source. It is important to note that x-ray measures a static distortion of the lattice. In the pressure region $p_1 < p < p_2$, the onset of magnetic order manifests itself in a sudden increase in orthorhombicity, thus enabling x-ray to be sensitive to the detection of long-range, static order. This comparison shows a very good agreement of the thermodynamic phase diagram with the one from x-ray scattering in terms of the transition temperatures $T_s$ and $T_M$. We only find a small discrepancy close to $p_2$. The analysis of x-ray data indicate that $p_2 \approx 1.6\,\mathrm{GPa}$. As a consequence, the data point at 1.7 GPa in Fig. S1 (b) is attributed to a single, first-order phase transition. In the present study, however, our analysis in the main text indicated that $p_2 = (1.88 \pm 0.1)\,\mathrm{GPa}$ and thus, slightly higher than the $p_2$ determined in x-ray studies. We attribute this slight discrepancy to experimental errors possibly arising from two different cells with different manometers and criterion for inferring pressure. In

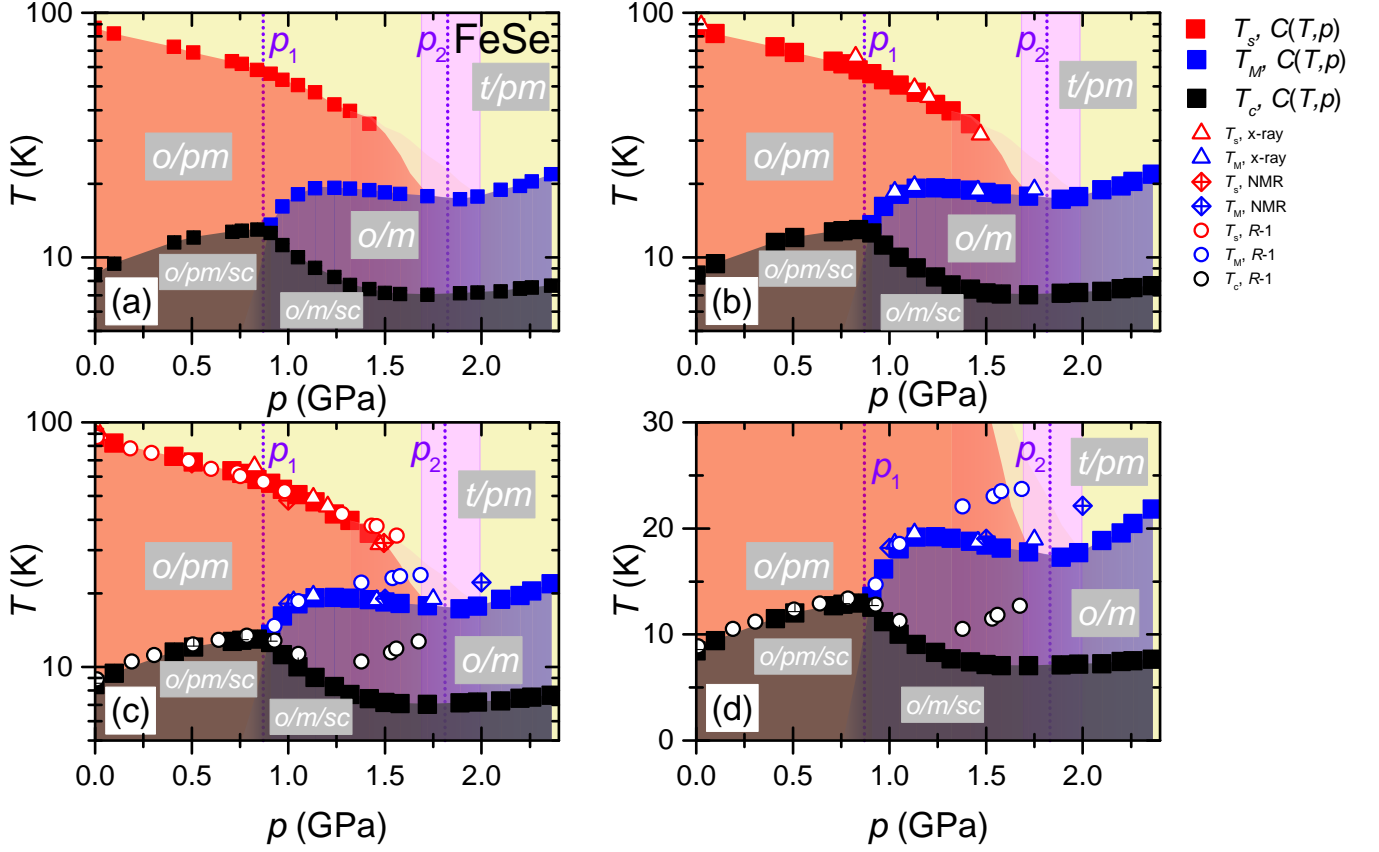

FIG. S1. (a) Temperature-pressure phase diagram of FeSe, as determined from the present specific heat study; (b) Comparison of phase diagram from specific heat with those from x-ray diffraction[9]; (c) Comparison of phase diagram from specific heat with data from x-ray diffraction[9], NMR[11] and resistance[10], all taken on crystals from the same source; (d) Blow-up of the phase diagram, shown in (c), on a linear scale. All labels are identical as the ones in Fig. 1 of the main manuscript.

Figs. S1 (c) and S1 (d), we add two additional $T$-$p$ phase diagrams from resistance[10] and NMR[11] measurements, both also taken on crystals from the same source. In this representation, and in particular in the blow-up in Fig. S1 (d) of the low-temperature phase diagram on a linear scale, the discrepancies of the magnetic transition temperatures $T_M$ and superconducting transition temperatures $T_c$ for $p > p_1$ become evident. This representation also highlights that our study actually identifies distinct break of slopes in $T_c(p)$ at the characteristic pressures $p_1$ and $p_2$. In contrast, the resistive $T_c(p)$ line shows only a pronounced change at $p_1$, but not at $p_2$ (see Fig. 4 in main text).

## B. Collection of raw data

In Fig. S2, we present the entire specific heat data set of this study, $C/T$ vs. $T$, ranging in pressure from $0\,\mathrm{GPa}$ to $2.36\,\mathrm{GPa}$ and in temperature from $5\,\mathrm{K}$ to $100\,\mathrm{K}$. We will discuss the salient features of these data sets in the subsections below, which is supplementary to the key information provided in the main manuscript.

## C. Structural transition

In Fig. S3 (e), we illustrate the determination of the structural transition temperature $T_s$ from our data sets (see Fig. S3 (a)-(d) and Fig. 1 of the main manuscript) by showing the derivative of the data sets. The jump-like change in $C/T$ at $T_s$ manifests itself in a distinct minimum in $\mathrm{d}(C/T)/\mathrm{d}T$ up to $1.32\,\mathrm{GPa}$. We assign the temperature, at

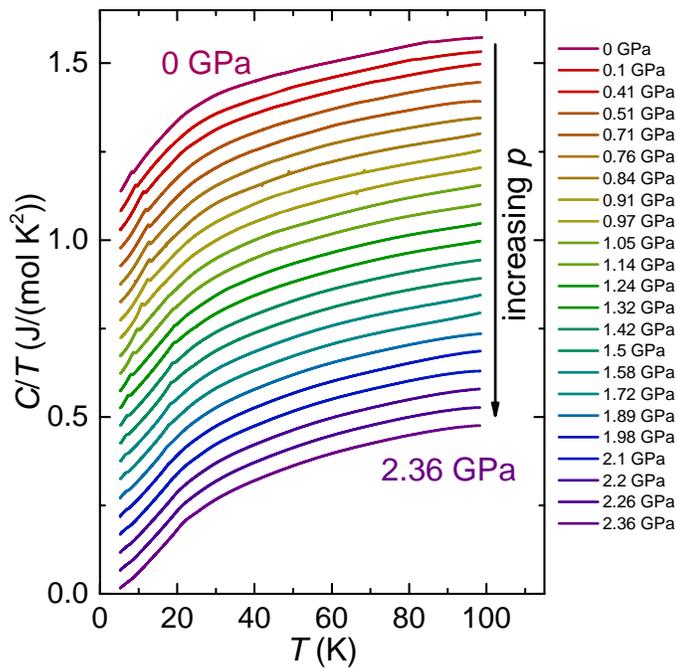

FIG. S2. Specific heat data, $C/T$ vs. $T$, on FeSe at different pressures from $0\,\mathrm{GPa}$ to $2.36\,\mathrm{GPa}$ across the full temperature range investigated. Data were successively offset by $0.05\ \mathrm{J/mol/K^2}$ for clarity.

which the minimum occurs, to $T_s$. Above $1.32\,\mathrm{GPa}$, the feature associated with $T_s$ becomes indiscernible (likely due to a combination of vanishing small entropy and strongly increasing slope of the phase transition line), as shown by adding the data set taken at $1.5\,\mathrm{GPa}$ as an example to Fig. S3 (e). In the latter data set, only the magnetic transition at $T_M \approx 18\,\mathrm{K}$ gives rise to a pronounced feature in $\mathrm{d}(C/T)/\mathrm{d}T$.

### D. Magnetic transition

#### 1. Criterion for determination of $T_M$ and background subtraction

In Fig. S4, we present the specific heat data, $C/T$, across the magnetic transition at $T_M$ at selected pressures (a,c,e), and the temperature derivatives of the respective data sets (b,d,f). The magnetic transition, which manifests itself in either a jump-like feature or a broad maximum in $C/T$ (depending on the pressure), shows up clearly as a step-like change in $\mathrm{d}(C/T)/\mathrm{d}T$ for $p \geq 0.97\,\mathrm{GPa}$. For $p = 0.91\,\mathrm{GPa}$ a $T_M$-associated feature can be identified (see Fig. S5), but it is at the limit of our ability to resolve. We assign the mid-point of the step-like change in $\mathrm{d}(C/T)/\mathrm{d}T$ to $T_M$. The so-derived $T_M$ values correspond well to the positions of the maxima in the background-corrected $\Delta C/T$ data, shown in Fig. 2 (a) of the main manuscript (see below for a discussion of the background correction).

Compared to resistance measurements under pressure, specific heat measurements give the opportunity to study whether any phase transition occurs below the superconducting transition temperature $T_c$. To discuss the possible extent of a magnetic transition below the superconducting one in FeSe at low pressures, we show in Fig. S5 the derivative of our specific heat data at $p = 0.84\,\mathrm{GPa}$, $0.91\,\mathrm{GPa}$ and $0.97\,\mathrm{GPa}$. At $0.97\,\mathrm{GPa}$, a clear feature of the magnetic transition in $\mathrm{d}(C/T)/\mathrm{d}T$ can be observed at $T_M \approx 15.9\,\mathrm{K} > T_c$. A similar feature, even though much smaller in size, can be observed at $T_M \approx 13.6\,\mathrm{K} > T_c$ at $0.91\,\mathrm{GPa}$. In both cases, no similar feature can be resolved below $T_c$ down to $6\,\mathrm{K}$. Similarly, no additional phase transition other than the superconducting one at $T < T_c$ as well as $T > T_c$ can be resolved in the data set, taken at $0.84\,\mathrm{GPa}$. Therefore, any feature of a magnetic transition for $T < T_c$ at all $p$, as well as for $T > T_c$ for $p \leq 0.84\,\mathrm{GPa}$, if exists, falls below our resolution.

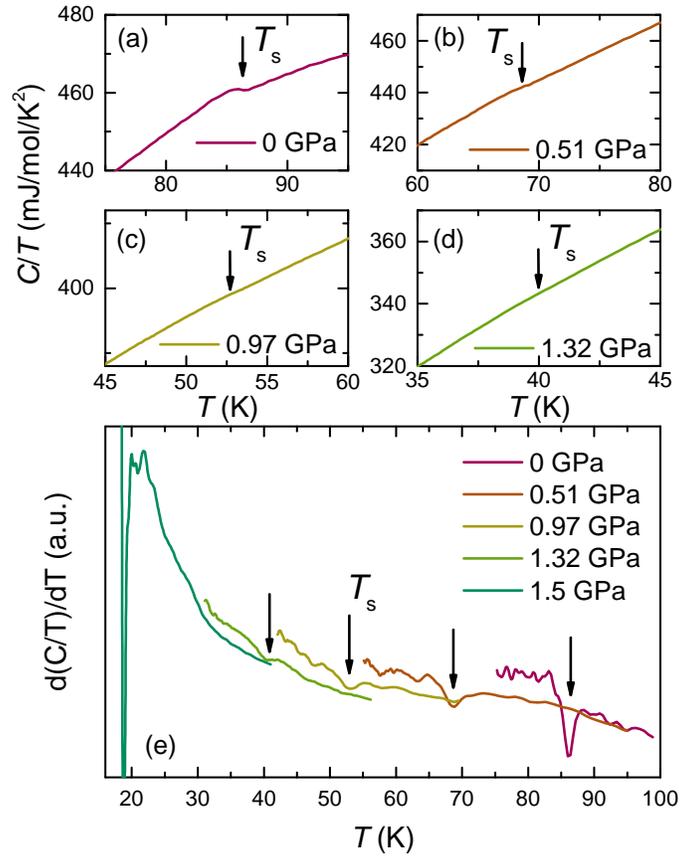

FIG. S3. Specific heat data of the structural phase transitions at $T_s$ in FeSe under pressure: (a)-(d) $C/T$ vs. $T$ at different pressures up to 1.32 GPa illustrating $T_s(p)$; (e) Derivative of the specific heat, d$(C/T)/$d$T$, vs. $T$ of the selected pressure data sets on FeSe, shown in (a)-(d). In addition, the derivative of the data set at 1.5 GPa is shown. The feature of the structural transition becomes indiscernible in this data set, the sharp feature at $T \approx 18$ K can be associated with the magnetic transition at $T_M$.

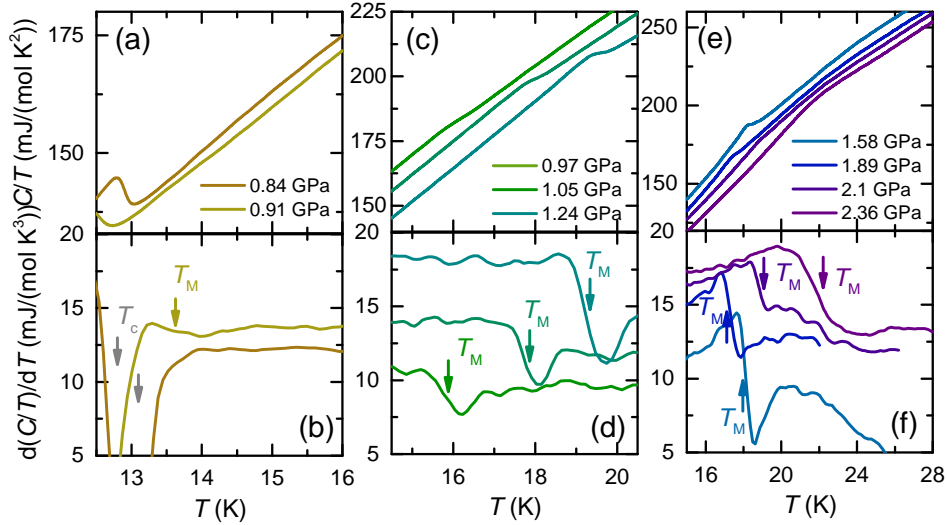

FIG. S4. Specific heat data, $C/T$, (top) and derivative of specific heat data, d$(C/T)/$d$T$, (bottom) on FeSe across the magnetic transition at $T_M$ at 0.84 GPa $\leq p \leq$ 0.91 GPa (a,b), 0.97 GPa $\leq p \leq$ 1.24 GPa (c,d) and 1.58 GPa $\leq p \leq$ 2.36 GPa (e,f). Arrows indicate the position of $T_M$ in the various data sets. Data in (b-d) and (f) were offset for clarity.

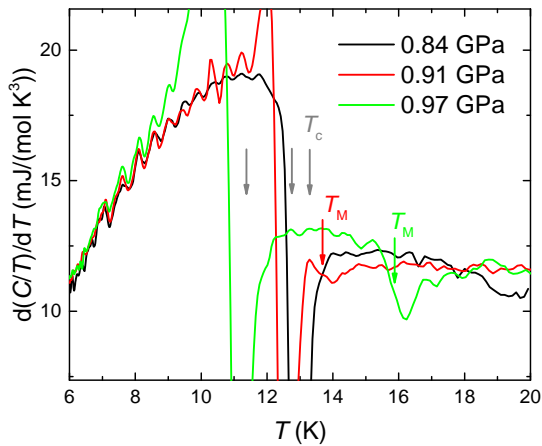

FIG. S5. Derivative of specific heat data, $d(C/T)/dT$ vs. $T$, on FeSe at 0.84 GPa, 0.91 GPa and 0.97 GPa. Grey arrows indicate the position of the superconducting transition at $T_c$ in the three data sets. Red and green arrows indicate the position of magnetic transition temperature at $T_M$, which is only discernible in data at $p \geq 0.91$ GPa above $T_c$, but not below $T_c$ in any of the data sets.

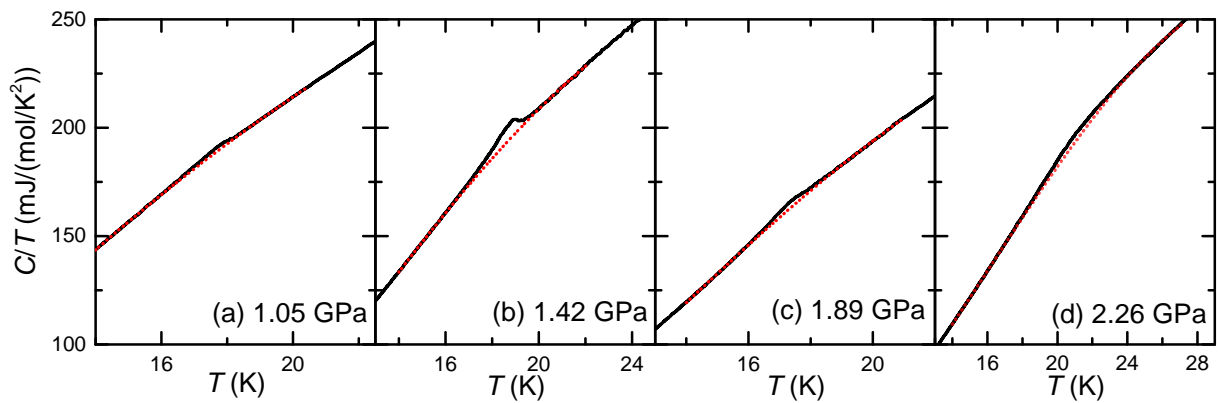

FIG. S6. Illustration of background subtraction to determine the anomalous contribution to the specific heat at the magnetic transition at selected pressures (a-d) in FeSe. The background (red dotted line) corresponds to a polynomial of order three which was fitted to the specific heat data (black line) at each pressure at $T < T_M$ and $T > T_M$ simultaneously.

To obtain the anomalous contribution to the specific heat at $T_M$, $\Delta C/T$, shown in Fig. 2(a) of the main manuscript, we have to estimate a background contribution due to the lack of reference background data when performing measurements under pressure. To this end, we approximate the background contribution to $C/T$ (see red dotted lines in Fig. S6) by fitting the $C/T$ data away from the phase transition temperature $T_M$ by a polynomial of the order of three. For each individual $p$ data set, we typically fit simultaneously in two temperature ranges, defined by $\approx (T_M - 3\,\mathrm{K}) \leq T \leq (T_M - 1\,\mathrm{K})$ and $\approx (T_M + 1\,\mathrm{K}) \leq T \leq (T_M + 3\,\mathrm{K})$.

### 2. Measurements of thermal hysteresis

In the main text, we use the size of the thermal hysteresis at $T_M$ to show that the transition changes its character from second order to first order at $p_2 \approx 1.65$ GPa. The $d(C/T)/dT$ data, which is used to determine this thermal hysteresis between warming and cooling, are shown in Fig. S7. These data were obtained using a slow heating/cooling rate of $\pm 0.25$ K/min to ensure thermal equilibrium. Using the midpoint of the step-like feature in $d(C/T)/dT$ (see *min, max* and *average* arrows in Fig. S7 (a)), we can determine $T_{M,warm}$ and $T_{M,cool}$, which are indicated by the dashed and dotted lines, and calculate the thermal hysteresis $\Delta T = T_{M,warm} - T_{M,cool}$. We find that the thermal hysteresis is almost constant below 1.63 GPa, and increases steadily above this pressure. Thus, the small hysteresis at

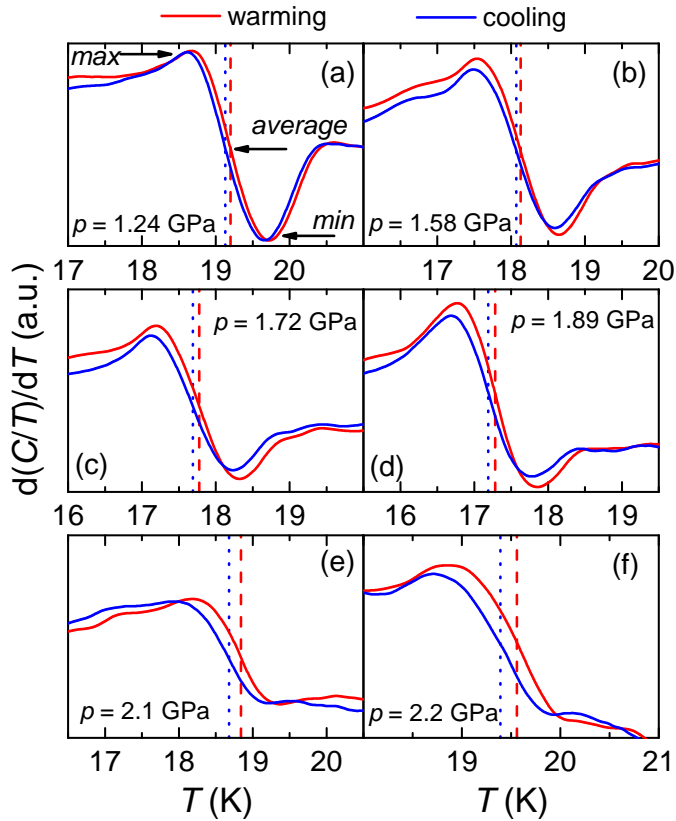

FIG. S7. Derivative of specific heat data, $\mathrm{d}(C/T)/\mathrm{d}T$ vs. $T$, taken upon warming (red) and cooling (blue) at selected pressures (a-f) across $p_2 = 1.65\,\mathrm{GPa}$ on FeSe. Dashed (dotted) lines indicate the position of the magnetic transition temperature $T_M$ upon warming (cooling). The arrows labeled with *max, average* and *min* are used to illustrate how the positions of the dotted lines were determined.

$p < 1.63\,\mathrm{GPa}$ is likely an instrumental hysteresis, inevitable in any low-temperature experiment, whereas the increase in $\Delta T$ at higher pressures reflects the hysteresis, related to the first-order character of the phase transition. We note that the herein observed hysteresis of $\Delta T < 100\,\mathrm{mK}$ at $1.7\,\mathrm{GPa}$ is fully consistent with the observations in Ref. 9, which claim that any hysteresis is smaller than their data point spacing of $200\,\mathrm{mK}$.

### E. Superconducting transition

#### 1. Raw data and background subtraction

Figure S8 shows the raw specific heat data, $C/T$, across the superconducting transition temperature $T_c$, at selected pressures in the three different pressure regimes (a) $p < p_1$, (b) $p_1 < p < p_2$ and (c) $p > p_2$, i.e., in the regimes in which superconductivity coexists with purely nematic order (a), nematic and magnetic order that occur at distinct ordering temperatures (b) and with strongly coupled magnetic and nematic order that occur simultaneously (c).

To obtain the estimate of the electronic specific heat, $\Delta C$, shown in the main manuscript, we have to subtract a background contribution from our bare $C/T$ data. Due to the lack of background information under pressure, we followed the standard procedure of subtracting a linear contribution from the $C/T$ data in a $T^2$ representation above $T_c$, as in a Fermi-liquid model $C = \gamma T + \beta T^3$ at low $T$, with the latter term describing the phononic specific heat. Examples of the linear fit in a $C/T$ vs. $T^2$ representation are shown in Fig. S9. However, this approach neglects the fact that superconducting fluctuations exist above $T_c$ (exceeding above $T_c$ by at least $10\,\mathrm{K}$, see main

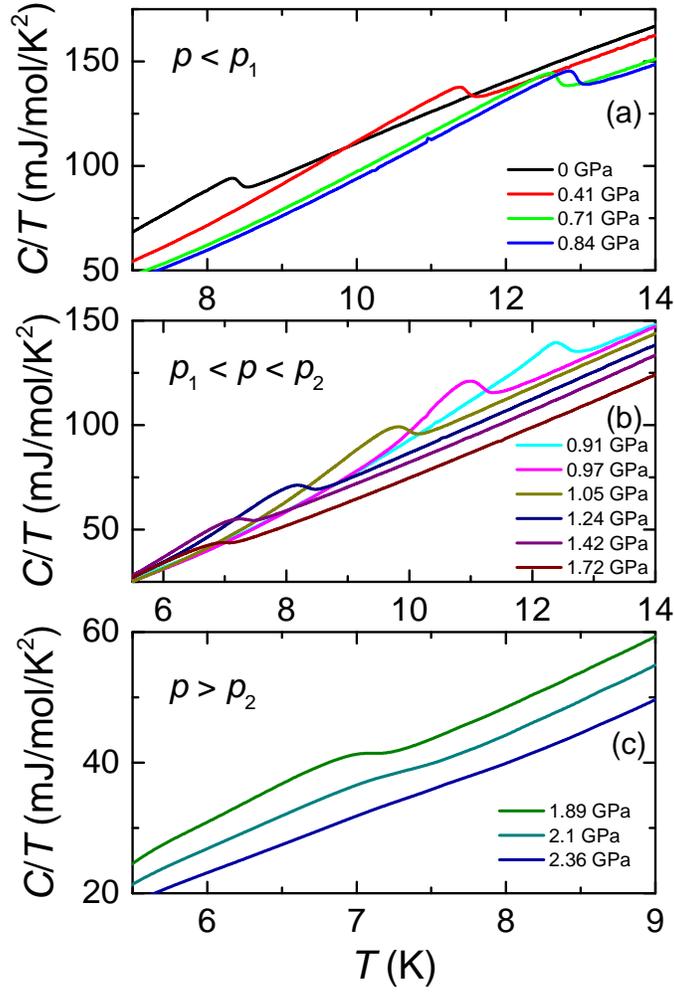

FIG. S8. Specific heat data, $C/T$, on FeSe across the superconducting transition at $T_c$ at selected pressures in the pressure ranges $p < p_1$ (a), $p_1 < p < p_2$ (b) and $p > p_2$ (c).

text) which likely give rise to an additional contribution to the specific heat at $T > T_c$. Unfortunately, due to the lack of background information under pressure, we cannot determine their contribution to the specific heat. As a consequence, the procedure shown here might yield an overestimate of background contribution and therefore an underestimate of the superconducting anomaly. Since the width of the specific heat transition at $T_c$ does not exceed 2 K in the investigated pressure range (see Fig. 3 (a) of the main manuscript), we nevertheless believe that we are able to determine a significant portion of $\Delta C_{sc}$ and therefore consider the suppression of $\Delta C_{sc}$ with $p$ at $p > p_1$ reliable.

### 2. Quantification of superconducting transition width

In the main text, we presented in Fig. 3 (d) the evolution of the width of the superconducting transition with pressure. To obtain this quantity, we used the following procedure, depicted in Fig. S10. In a clean, BCS superconductor, the specific heat would exhibit a sharp jump at $T_c$, thus giving rise to an infinite sharp peak in the derivative $d(\Delta C/T)/dT$. In any real system, however, the jump in the specific heat will be broadened and, as a consequence, the peak in the derivative will exhibit a finite width. The causes of this broadening can be multifold: small amounts of disorder, small pressure inhomogeneities, or also the presence of some critical fluctuations above $T_c$ beyond the BCS mean-field description, as intensively discussed for the present case in the main text, etc.. One possible way to quantify this broadening is to quantify the finite width of the peak in the derivative by determining its full width

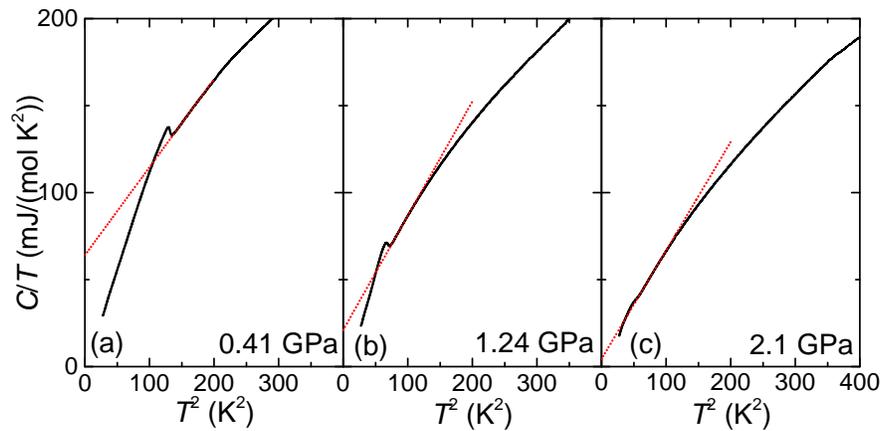

FIG. S9. Illustration of background subtraction to determine the electronic contribution to the specific heat at low temperatures at selected pressures (a-c) in FeSe. The background (red dotted line) corresponds to a linear fit to the specific heat data (black line), plotted in a $C/T$ vs. $T^2$ representation, above $T_c$. The strong change of background contribution results from changes of the electronic and phononic contributions. The former one is likely to change significantly due to the presence of magnetism above $T_c$ at $p > p_1 \approx 0.9\,\mathrm{GPa}$.

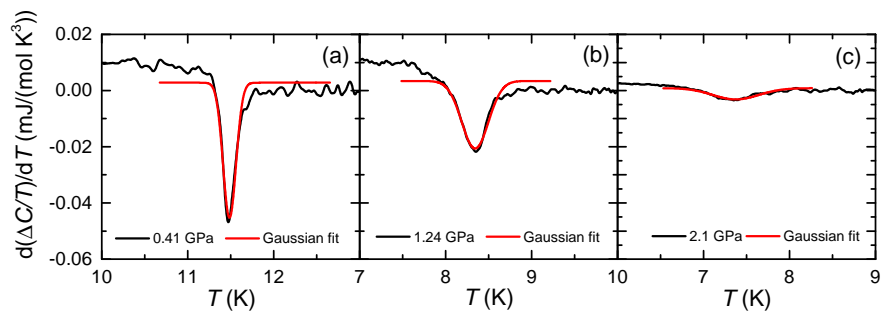

FIG. S10. Illustration of the procedure to determine the width of the superconducting transition at selected pressures (a-c). The black line corresponds to the temperature derivative of the $\Delta C/T$ data, the red line to a Gaussian fit to the experimental data.

at half maximum. Thus, we fitted the peak in our $\mathrm{d}(\Delta C/T)/\mathrm{d}T$ by a Gaussian peak function and compiled the full width at half maximum of each of these fits at the respective pressure in Fig. 3 (d) of the main manuscript.

The sudden increase of broadening of the specific heat features at the superconducting transition at $p_1 = 0.91\,\mathrm{GPa}$, i.e. at the onset of magnetism, is already apparent from the raw data. To show this, we present in Fig. S11 data of the specific heat at different pressures as a function of $T - T_c$. At each pressure, the specific heat was normalized to its peak value and therefore is labeled with $\Delta C_{normalized}$. While the two data sets at $0\,\mathrm{GPa}$ and $0.84\,\mathrm{GPa}$ were taken in the non-magnetic state, the data sets at $0.91\,\mathrm{GPa}$ and $1.42\,\mathrm{GPa}$ were taken in the magnetic state. It is obvious, that the $\Delta C_{normalized}$ feature is distinctly broader for $0.91\,\mathrm{GPa}$ and $1.42\,\mathrm{GPa}$ than for the lower pressure data sets.

In Fig. S12, we compare the pressure evolution of superconducting transition width with the one of the magnetic transition. We find that the superconducting as well as the magnetic transition width exhibit a rapid increase above $\approx 2\,\mathrm{GPa}$, i.e., very close to $p_2$. However, we would like to point out that the discrepancies between the determined transition temperatures $T_c$ and $T_M$ of different studies at $p > 2\,\mathrm{GPa}$ are distinctly larger than the transition widths determined here. Thus, this observation does not significantly affect the main message of this paper.

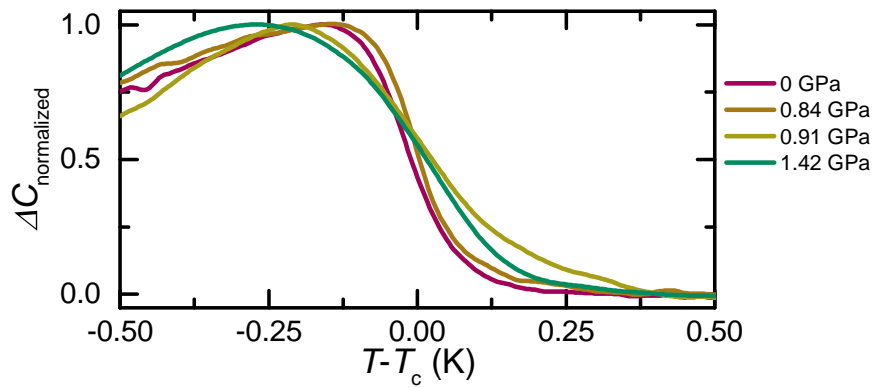

FIG. S11. Specific heat around $T_c$, normalized to its maximum value, $\Delta C_{normalized}$, vs. $T - T_c$ for $p < p_1$ (0 GPa and 0.84 GPa) and $p_1 < p < p_2$ (0.91 GPa and 1.42 GPa).

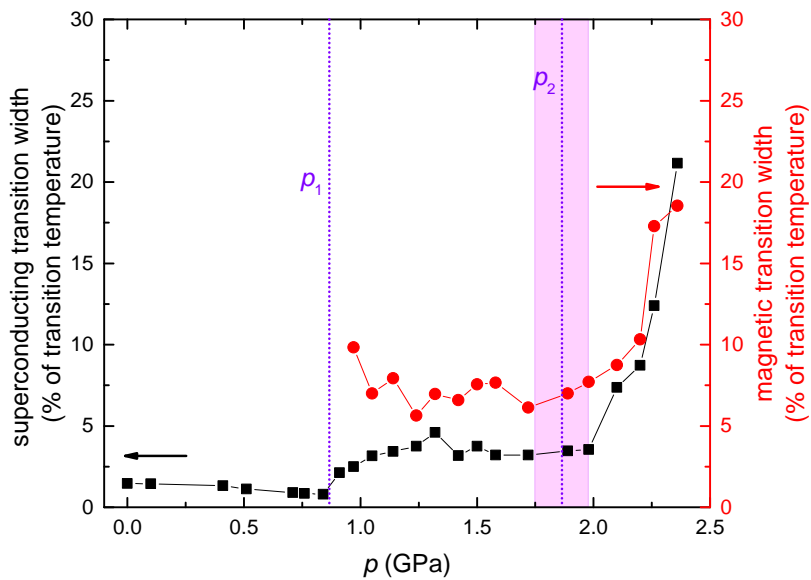

FIG. S12. Pressure evolution of superconducting transition width (black squares, left axis) and magnetic transition width (red circles, right axis), both given in percent of the respective transition temperature.

## F. Comparison of specific heat data with resistance, magnetization and x-ray data

In this section, we compare our temperature-dependent specific heat data around the superconducting and magnetic transition at selected pressures with those of resistance[10], dc magnetization[12], ac susceptibility[13] and x-ray[9]. While the good agreement of specific heat and x-ray data around the magnetic transition was discussed in great detail, we aim to discuss here the anomalous behavior of the superconducting transition in the magnetically-ordered state, which was initiated by earlier studies and is the focus of the present work, in more detail. So far, the anomalous behavior at $p > p_1$ was mainly associated with the following two aspects: (i) the resistive transition is significantly broadened compared to the $p < p_1$ range, but becomes sharp once the magnetic-nematic state is suppressed at $\approx$ 6 GPa[14] and (ii) the onset of diamagnetism under pressure in the presence of magnetism typically does not coincide with the temperature at which resistance reaches zero[13]. In the present work, we established yet another peculiarity in the magnetically-ordered state, namely that the bulk transition temperature at $T_{c,C}$ determined from specific heat is even lower than the onset of diamagnetism ($T_{c,M}$ for dc magnetization measurements and $T_{c,\chi}$ for ac susceptibility) and the temperature at which resistance reaches zero. This is clearly illustrated in Fig. S13 (a-d) at $p \approx 1.6$ GPa and (e-g) at $p \approx 1.7$ GPa. We find that $T_{c,C} < T_{c,M} < T_{c,R}$ at 1.58 GPa, and $T_{c,C} < T_{c,\chi}$ at 1.7 GPa. In this discussion, it is important to note that resistance and specific heat data were obtained on crystals from the same source in the same pressure cell environment, whereas magnetization and ac susceptibility data were taken in different conditions. The

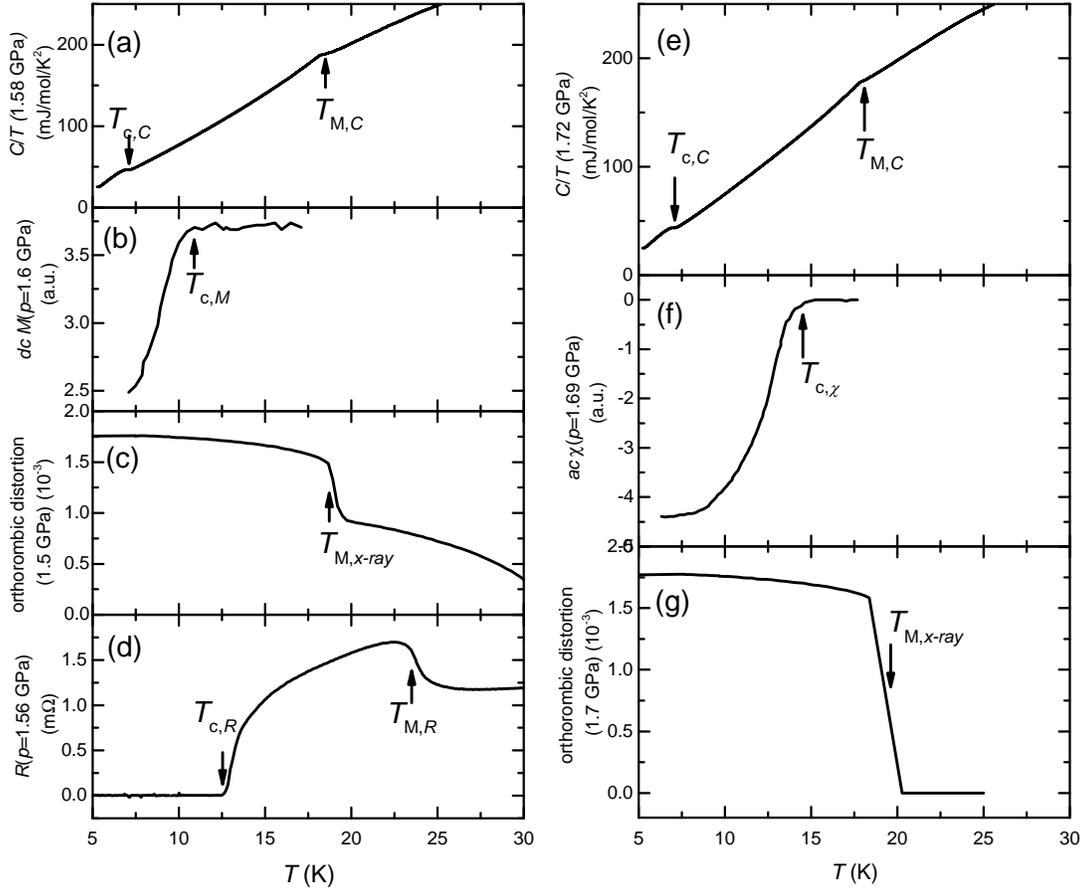

FIG. S13. (a-d) Comparison of specific heat data (a) on FeSe at 1.58 GPa to dc magnetization data at 1.6 GPa (Ref. [12]) (b), to x-ray data at 1.5 GPa (Ref. [9]) (c) and (d) resistance data at 1.56 GPa (Ref. [10]) and ; (e-g) Comparison of specific heat data (e) on FeSe at 1.72 GPa to ac susceptibility data at 1.6 GPa (Ref. [13]) (f) and to x-ray data at 1.7 GPa (Ref. [9]) (g).

comparisons in Fig. S13 show that the resistive $T_{c,R}$ is actually not accompanied by any significant shielding fraction. It is well known, though, that resistance measurements in superconductors might be fooled by a short circuit in the sample, induced by a tiny volume fraction. It is more peculiar that also the onset of diamagnetism at $T_{c,M}$ and $T_{c,\chi}$ does not coincide with the specific heat transition at $T_{c,C}$. Even further, it appears that $T_{c,C}$ rather coincides with the saturation of the diamagnetism (see Fig.S13 (b)). The survival of diamagnetism above $T_c$ is observed less commonly in superconductors, but has recently been discussed in FeSe even at ambient pressure as a consequence of superconducting fluctuations surviving far above $T_c$[15] in the crossover between BCS and BEC superconductivity. Even though this scenario at ambient pressure is debated at present[16] and we show that $T_{c,C} \simeq T_{c,M} \simeq T_{c,R}$ at low pressures ($p < p_1$), our results indicate at the same time that the onset of magnetic order, for $p > p_1$, in FeSe triggers a much more pronounced difference between onset of diamagnetism and bulk transition. It will be therefore important in the future to identify why the onset of bulk magnetic order goes along with diamagnetism surviving far above $T_{c,C}$ in FeSe under pressure.

### G. Further comments on the competing nature of superconductivity and magnetism in FeSe

The competition of superconductivity and magnetism in iron-based superconductors is often studied by microscopic probes, which measure the strength of the magnetic hyperfine field[17–19]. In this situation, the onset of superconductivity usually results in a decrease of the magnetic hyperfine field. However, whether this decrease is pronounced or not, depends typically on the relative strength of superconductivity to magnetism (in terms of transition temperatures and respective densities of states), and therefore the decrease is sometimes lower than the resolution limit of the respective technique.

Based on our thermodynamic analysis of the specific heat jump size in FeSe, we suggest that magnetism and superconductivity compete with each other, whenever magnetism is present. In this pressure region ($p > p_1$), we propose that magnetism becomes strengthened compared to superconductivity. As a consequence, a decrease of the magnetic hyperfine field might only be observable at lower pressures, close to but greater than $p_1$. Indeed, $\mu$SR measurements[20] were utilized to evaluate the magnetic hyperfine field and demonstrated that at low pressures ($p \approx 1$ GPa) the hyperfine field is decreased upon entering the superconducting state. This experimental observation therefore strongly supports our proposal. In more detail, in these $\mu$SR works, it was shown that below $T_c$ the ordered moment, as well as the volume fraction, decrease. Similar observations were made more recently for the iron-based superconductor NaFe$_{1-x}$Ni$_x$As[21]. In general, the competition in itinerant systems is captured by a Ginzburg-Landau approach, in which a parameter $g$ measures the degree of competition of magnetism and superconductivity. In the case of small $g$, a microscopic coexistence can be established, which in turn should result, in a homogeneous system, only in a reduction of the moment size, but not the volume fraction. In contrast, for large $g$, the two orders are macroscopically phase separated, which then should result in a reduction of the volume fraction, but not the moment. Based on these general arguments, it was concluded[21] that FeSe, just as NaFe$_{1-x}$Ni$_x$As, is at the verge of phase separation and microscopic coexistence. The specific heat anomaly as such is a measure of the entropy change, which is a product of both, the volume and moment size. Consequently, our data is consistent with microscopic or macroscopic phase coexistence, but would also be compatible with a scenario where FeSe is at the verge between these two extreme limits.

In terms of the relative evolution of transition temperatures $T_c$ and $T_M$, we pointed out in the main manuscript that a competition does not necessarily imply that the respective transition temperatures behave in opposite manners upon external tuning, i.e., the slopes of the transition temperatures with respect to the tuning parameter do not have to be of opposite sign. We demonstrated that this is indeed the case for FeSe, where both transition temperatures actually increase, however with distinct slopes. We rationalized this finding with the help of the model of Ref. 22. We would like to add here that FeSe is not unique among the iron-based superconductors in this respect. Mössbauer measurements on the recently discovered family CaK(Fe$_{1-x}$Ni$_x$)$_4$As$_4$[17] demonstrated that superconductivity and magnetism compete. Pressure studies[23] on magnetically-ordered members of this family actually demonstrated that both $T_M$ and $T_c$ decrease, with $|\mathrm{d}T_M/\mathrm{d}p| < |\mathrm{d}T_c/\mathrm{d}p|$. Following our ideas on FeSe, this behavior of CaK(Fe$_{1-x}$Ni$_x$)$_4$As$_4$ this behavior is also fully consistent with the competing nature of both orders.

---

[1] A. E. Böhmer, V. Taufour, W. E. Straszheim, T. Wolf, and P. C. Canfield, Phys. Rev. B **94**, 024526 (2016).

[2] E. Gati, G. Drachuck, L. Xiang, L.-L. Wang, S. L. Bud'ko, and P. C. Canfield, Review of Scientific Instruments **90**, 023911 (2019).

[3] F. Hardy, M. He, L. Wang, T. Wolf, P. Schweiss, M. Merz, M. Barth, P. Adelmann, R. Eder, A.-A. Haghighirad, et al., Phys. Rev. B **99**, 035157 (2019).

[4] A. Muratov, A. Sadakov, S. Gavrilkin, A. Prishchepa, G. Epifanova, D. Chareev, and V. Pudalov, Physica B: Condensed Matter **536**, 785 (2018).

[5] Y. Sun, S. Kittaka, S. Nakamura, T. Sakakibara, K. Irie, T. Nomoto, K. Machida, J. Chen, and T. Tamegai, Phys. Rev. B **96**, 220505 (2017).

[6] S. K. Kim, M. S. Torikachvili, E. Colombier, A. Thaler, S. L. Bud'ko, and P. C. Canfield, Phys. Rev. B **84**, 134525 (2011).

[7] M. S. Torikachvili, S. K. Kim, E. Colombier, S. L. Bud'ko, and P. C. Canfield, Review of Scientific Instruments **86**, 123904 (2015).

[8] B. Bireckoven and J. Wittig, Journal of Physics E: Scientific Instruments **21**, 841 (1988).

[9] K. Kothapalli, A. E. Böhmer, W. T. Jayasekara, B. G. Ueland, P. Das, A. Sapkota, V. Taufour, Y. Xiao, E. Alp, S. L. Bud'ko, et al., Nat. Commun. **7**, 12728 (2016).

[10] U. S. Kaluarachchi, V. Taufour, A. E. Böhmer, M. A. Tanatar, S. L. Bud'ko, V. G. Kogan, R. Prozorov, and P. C. Canfield, Phys. Rev. B **93**, 064503 (2016).

[11] P. Wiecki, M. Nandi, A. E. Böhmer, S. L. Bud'ko, P. C. Canfield, and Y. Furukawa, Phys. Rev. B **96**, 180502 (2017).

[12] K. Miyoshi, K. Morishita, E. Mutou, M. Kondo, O. Seida, K. Fujiwara, J. Takeuchi, and S. Nishigori, Journal of the Physical Society of Japan **83**, 013702 (2014).

[13] T. Terashima, N. Kikugawa, S. Kasahara, T. Watashige, T. Shibauchi, Y. Matsuda, T. Wolf, A. E. Böhmer, F. Hardy, C. Meingast, et al., Journal of the Physical Society of Japan **84**, 063701 (2015).

[14] J. P. Sun, K. Matsuura, G. Z. Ye, Y. Mizukami, M. Shimozawa, K. Matsubayashi, M. Yamashita, T. Watashige, S. Kasahara, Y. Matsuda, et al., Nat. Commun. **7**, 12146 (2016).

[15] S. Kasahara, T. Yamashita, A. Shi, R. Kobayashi, Y. Shimoyama, T. Watashige, K. Ishida, T. Terashima, T. Wolf, F. Hardy, et al., Nature Communications **7**, 12843 (2016).

[16] H. Yang, G. Chen, X. Zhu, J. Xing, and H.-H. Wen, Phys. Rev. B **96**, 064501 (2017).

[17] S. L. Bud'ko, V. G. Kogan, R. Prozorov, W. R. Meier, M. Xu, and P. C. Canfield, Phys. Rev. B **98**, 144520 (2018).

[18] W. R. Meier, Q.-P. Ding, A. Kreyssig, S. L. Bud'ko, A. Sapkota, K. Kothapalli, V. Borisov, R. Valentí, C. D. Batista, P. P.

Orth, et al., npj Quantum Materials **3**, 5 (2018).

[19] A. Kreyssig, J. M. Wilde, A. E. Böhmer, W. Tian, W. R. Meier, B. Li, B. G. Ueland, M. Xu, S. L. Bud'ko, P. C. Canfield, et al., Phys. Rev. B **97**, 224521 (2018).

[20] M. Bendele, A. Ichsanow, Y. Pashkevich, L. Keller, T. Strässle, A. Gusev, E. Pomjakushina, K. Conder, R. Khasanov, and H. Keller, Phys. Rev. B **85**, 064517 (2012).

[21] S. C. Cheung, Z. Guguchia, B. A. Frandsen, Z. Gong, K. Yamakawa, D. E. Almeida, I. J. Onuorah, P. Bonfá, E. Miranda, W. Wang, et al., Phys. Rev. B **97**, 224508 (2018).

[22] K. Machida, Journal of the Physical Society of Japan **50**, 2195 (1981).

[23] L. Xiang, W. R. Meier, M. Xu, U. S. Kaluarachchi, S. L. Bud'ko, and P. C. Canfield, Phys. Rev. B **97**, 174517 (2018).



\end{document}